\documentclass[a4paper,12pt]{article}
\usepackage{graphicx,bbm}
\usepackage{amsmath,amsfonts,amssymb}
\usepackage{fleqn} 
\usepackage{mathrsfs}
\usepackage{epsfig}\parskip 5pt plus 1pt
\usepackage[footnotesize]{caption}

\newcommand{\SuperField}[1]{\hat{#1}}

\def\Ng{j}

\def\SU{\text{SU}}

\def\beq{\begin{equation}}
\def\eeq{\end{equation}}
\def\bea{\begin{eqnarray}}
\def\eea{\end{eqnarray}}

\def\eps{\epsilon}

\newcommand{\gsim}{\lower.7ex\hbox{$\;\stackrel{\textstyle>}{\sim}\;$}}
\newcommand{\lsim}{\lower.7ex\hbox{$\;\stackrel{\textstyle<}{\sim}\;$}}

\def\<{\left\langle}
\def\>{\right\rangle}
\def\ChargeC{\mathrm{C}} 
\def\chargec{\mathrm{C}}

\begin{document}

\bibliographystyle{OurBibTeX}

\begin{titlepage}

\vspace*{-15mm}
\begin{flushright}
CERN-PH-TH/2006-174\\ 
FTUAM 06-14\\ 
IFT-UAM/CSIC 06-43\\
SHEP-06-25\\
hep-ph/0609038
\end{flushright}
\vspace*{3mm}

\begin{center}
{
\sffamily
\Large
Flavour-Dependent Leptogenesis with Sequential Dominance}
\\[8mm]
S.~Antusch$^{\star}$
\footnote{E-mail: \texttt{antusch@delta.ft.uam.es}},
S.~F.~King$^{\dagger}$
\footnote{E-mail: \texttt{sfk@hep.phys.soton.ac.uk}}, 
A.~Riotto$^{\ddagger,\diamond}$
\footnote{E-mail: \texttt{antonio.riotto@pd.infn.it}}
\\[1mm]

\end{center}
\vspace*{0.50cm}
\centerline{$^{\star}$ \it Departamento de Fisica Te\'orica C-XI and Instituto
del Fisica Te\'orica C-XVI,}
\centerline{\it 
Universidad Aut\'onoma de Madrid, Cantoblanco, E-28049 Madrid, Spain}
\centerline{$^\dagger$ \it School of Physics and Astronomy,}
\centerline{\it  
University of Southampton,
Sounthampton, SO17 1BJ, UK}
\centerline{$^\ddagger$ \it CERN, Theory Division, Geneva 23, Ch-1211, Switzerland}
\centerline{$^\diamond$ 
\it INFN, Sezione di Padova, via Marzolo 8, I-35131 Padova,
Italy}
\vspace*{0.50cm}

\begin{abstract}

\noindent 
We study thermal leptogenesis in classes of neutrino mass models based
on the seesaw mechanism with three right-handed neutrinos and 
sequential right-handed neutrino dominance. 
The flavour-dependent Boltzmann equations are solved appropriate to both the Standard Model and the Minimal Supersymmetric Standard Model. 
Within these classes of models we investigate 
constraints and expectations on the individual decay asymmetries and 
washout parameters from the present data on neutrino masses and mixings.
In many cases of physical interest
flavour effects are shown to have important consequences for the
estimation of the produced baryon asymmetry in leptogenesis.
We also establish and analyze the link between the leptonic CP violating 
phase $\delta$, observable in neutrino oscillations, and the CP violation
required for leptogenesis, where flavour-dependent effects
have a significant effect. In general our results show 
that flavour-dependent effects cannot be ignored when dealing with three
right-handed neutrino models.

\end{abstract}

\end{titlepage}
\newpage
\setcounter{footnote}{0}

\section{Introduction}
Thermal leptogenesis \cite{FY} is an attractive and minimal mechanism to
generate the baryon asymmetry of the Universe
$n_\mathrm{B} /n_\gamma 
\,\approx\, (6.10\,\pm\,0.21)\,\times\,10^{-10}$,
or, normalised to the entropy density, 
$Y_B\,\approx\, (0.87\,\pm\,0.03)\,\times\,10^{-10}$ \cite{Spergel:2006hy}.
A lepton asymmetry is dynamically generated
and then  converted into a baryon asymmetry
due to $(B+L)$-violating sphaleron interactions \cite{kuzmin}
which exist in the Standard Model (SM) and its minimal supersymmetric extension, the MSSM.
Leptogenesis  can be implemented within  the 
``Seesaw''(type I) model 
\cite{seesaw}, consisting
of the SM (MSSM)
plus  three   right-handed (RH) Majorana neutrinos (and their superpartners)
with a hierarchical spectrum. In the simplest case,  the 
lightest of the RH neutrinos
is produced by thermal scattering after inflation,  and subsequently
decays out-of-equilibrium in a lepton number and CP-violating way, 
thus satisfying Sakharov's constraints  \cite{sakharov}.

The asymmetry
is commonly calculated by solving a Boltzmann equation for the
total lepton asymmetry (flavour-independent approximation) 
and for the abundance of the 
lightest RH neutrino \cite{lept,ogen,work}. However, this 
flavour-independent treatment is
rigorously correct only when the interactions mediated by 
charged lepton Yukawa couplings are out of equilibrium.
The impact of  flavour in thermal leptogenesis has been first addressed in
Ref.~\cite{barbieri} and then 
studied in  detail 
\cite{davidsonetal,nardietal,Abada:2006ea}, including the
quantum oscillations/correlations of the asymmetries in lepton flavour space 
\cite{davidsonetal}. It was shown that 
the Boltzmann equations describing the asymmetries in flavour space 
have  additional terms which can  significantly affect the result for the
final baryon asymmetry.
This is because
leptogenesis involves  the  production and 
destruction of right-handed neutrinos, 
and  of a  lepton asymmetry  that is distributed among
{\it distinguishable} flavours.  The processes which
wash out lepton number
are {\it flavour-dependent}, e.g., the inverse decays
from electrons can destroy the lepton asymmetry carried by,
and only by,  the electrons.
The  asymmetries  in each flavour
are  therefore washed out differently,
and will appear with different weights in the final formula
for the baryon asymmetry.  This is physically inequivalent
to the treatment of washout in the flavour-independent approximation,
where indistinguishable 
leptons propagate between decays and inverse decays,
so inverse decays  from all flavours are taken to wash out asymmetries
in any flavour. Flavour-dependent
effects in leptogenesis 
have been studied in the two right-handed neutrino model 
\cite{Abada:2006ea}, where they have been shown to be 
relevant. 
It remains to be seen how important these effects are in more general 
neutrino mass models with three right-handed neutrinos.

The latest experimental data on neutrino oscillations
is consistent with (approximate) tri-bimaximal
mixing \cite{tribi}. The fact that the tri-bimaximal neutrino
mixing matrix involves
square roots of simple ratios motivates models in which the mixing
angles are independent of the mass eigenvalues. One such class of
models are seesaw models with sequential dominance (SD) of
right-handed neutrinos \cite{King:1998jw}.
In SD, a neutrino mass hierarchy is shown to
result from
having one of the right-handed neutrinos give the dominant
contribution to the
seesaw mechanism, while a second right-handed neutrino gives the
leading sub-dominant contribution, leading to a neutrino mass
matrix with naturally small determinant.
In a basis where the right-handed neutrino mass matrix is
diagonal, the atmospheric and solar neutrino mixing angles are determined in
terms of ratios
of Yukawa couplings involving the dominant and subdominant
right-handed neutrinos, respectively. If these Yukawa couplings
are simply related in some way, then it is possible for simple
neutrino mixing angle relations, such as appear in tri-bimaximal
neutrino mixing, to emerge in a simple and natural way,
independently of the neutrino mass eigenvalues.
If SD is combined with
vacuum alignment in flavour models then tri-bimaximal
neutrino mixing can be readily achieved 
\cite{King:2005bj,deMedeirosVarzielas:2005ax}.
The SD mechanism has been widely applied to a large variety 
of unified flavour models 
involving three right-handed neutrinos \cite{Antusch:2004gf},
and it is therefore of interest to see how important the effects
of flavour-dependent leptogenesis are in this more general framework.

In this paper we study thermal leptogenesis in classes of neutrino mass models 
based on the seesaw mechanism with sequential
dominance, taking into account lepton flavour in the Boltzmann equations. 
We generalize the issue of 
including flavour effects in the Boltzmann equations to the 
supersymmetric case. Within this class of model
we investigate constraints and expectations on the individual decay
asymmetries and washout parameters from the present data on neutrino masses 
and mixings.
Flavour effects are shown to have important consequences for the estimation 
of the produced baryon asymmetry in this class of models.
Flavour-independent leptogenesis has previously been considered
for sequential dominance models in \cite{Hirsch:2001dg}.
The present analysis clearly goes well beyond the previous
analysis by considering the important flavour-dependent effects.
Also the link between the leptogenesis phase and the MNS
neutrino oscillation phase has been explored in sequential dominance in
\cite{King:2002qh}. Here we shall revisit this link in the light
of flavour effects and obtain new links between the cosmology and
neutrino oscillation physics.

The remainder of the paper is set out as follows: In
Sec.~\ref{Sec:FML}, we review the treatment of flavour in the
Boltzmann equations and generalize it to the case of the
MSSM. Sec.~\ref{SD} contains a review of sequential dominance (SD) in
the seesaw mechanism. In Sec.~\ref{Sec:FlavourLGinSD}, we investigate
leptogenesis in neutrino mass models with SD, taking into account
lepton flavour in the Boltzmann equations. Generic properties of the
decay asymmetries and washout parameters are derived in
Sec.~\ref{Sec:DecayAndWashout} and the link between the leptonic CP
violating phase $\delta$, observable in neutrino oscillations, and the
CP violation required for leptogenesis is established in
Sec.~\ref{Sec:MNSLGlink}. Sec.~\ref{examples} contains examples to
illustrate our results. In Sec.~\ref{concl} we conclude.

\section{Flavour Matters in Leptogenesis}\label{Sec:FML}
\subsection{Temperatures where Flavour Matters}
In the SM extended by right-handed (singlet) neutrinos $N_i$ ($i=1,2,3$) with Majorana masses $M_i$, the additional terms of the Lagrangian are given by
\begin{eqnarray}\label{Eq:L}
 \mathcal{L} & = &
- (\lambda_\nu)_{\alpha i}({\ell}^{\alpha}\cdot
 H)\,  N^{i} 
 - \frac{1}{2} \overline{N^{i}} (M_\mathrm{RR})_{ij}
 N^{\ChargeC j}\;+\;\text{H.c.}\;,
\end{eqnarray}
where the dot indicates the $\SU (2)_\mathrm{L}$-invariant product, 
$(\ell^{\alpha}\cdot H) := \ell_a^{\alpha}(i\tau_2)^{ab} ({H})_b$, 
with $\tau_A$ $(A\in \{1,2,3\})$ being the Pauli matrices. 
$\ell_\alpha$ ($\alpha=e,\mu,\tau$) are the lepton 
SU(2)$_\mathrm{L}$-doublets and $H$ is the Higgs field which develops a vacuum expectation value (vev) of $\<H^0\> \equiv v_\mathrm{u} = 175$ GeV in its neutral component after electroweak symmetry breaking. We will work in a basis where the charged lepton Yukawa matrix and the mass matrix of the right-handed neutrinos are diagonal, i.e.\ $\lambda_e = \mbox{diag}(y_e,y_\mu,y_\tau)$ and $M_\mathrm{RR}= \mbox{diag}(M_1,M_2,M_3)$, respectively. We will assume a hierarchical spectrum of right-handed neutrino masses, $M_1 \ll M_2 \ll M_3$, in the following. 

In the SM, the flavour-independent formulae
to describe leptogenesis  are only appropriate when 
the dynamics takes place at 
temperatures larger than about $10^{12}$ GeV, before the charged lepton Yukawa couplings
come into equilibrium, estimating the interaction rate for a Yukawa coupling $y_\alpha$ as $\Gamma_\tau \approx 5 \times 10^{-3} \,y_\alpha^2 \,T$ \cite{Cline:1993bd}.  However, if leptogenesis occurs at smaller 
temperatures $T\sim M_1$, where $M_1$ is the mass of the lightest RH neutrino,
then one has to distinguish two possible cases. If 
 $10^5 \: \mbox{GeV} \ll M_1 \ll 10^{9} \: \mbox{GeV}$,  
then charged $\mu$ and $\tau$ Yukawa couplings are in thermal equilibrium 
and all flavours in the Boltzmann equations are to be 
treated separately. For 
$10^9 \: \mbox{GeV} \ll M_1 \ll 10^{12} \: \mbox{GeV}$, only the $\tau$ Yukawa coupling is in equilibrium and is treated separately in the Boltzmann equations, while
the $e$ and $\mu$ flavours are indistinguishable.

In the MSSM, extended by singlet superfields $\SuperField{N}^\chargec_i$ ($i=1,2,3$) containing the right-handed neutrinos $N^i$ as fermionic components, we use a notation analogous to the SM. 
The additional terms of the superpotential are given by  
\begin{eqnarray}\label{Eq:W}
 \mathcal{W} & = &
(\lambda_\nu)_{\alpha \Ng}(\SuperField{\ell}^{\alpha}\cdot
 \SuperField{H}_\mathrm{u})\,  \SuperField{N}^{\chargec j} 
 + \frac{1}{2} \SuperField{N}^{\chargec i} (M_\mathrm{RR})_{ij}
 \SuperField{N}^{\chargec j}\;,
\end{eqnarray}
where hats denote superfields. In the MSSM, the vev of the Higgs field $H_\mathrm{u}$, which couples to the right-handed neutrinos, is given by $\<H_\mathrm{u}^0\> \equiv v_\mathrm{u} = \sin (\beta) \times 175$ GeV, with $\tan \beta$ defined as usual as the ratio of the vevs of the Higgs fields which couple to up-type quarks (and right-handed neutrinos) and down-type quarks (and charged leptons). 

In the MSSM, the flavour-independent formulae can only be applied for temperatures larger than $(1+\tan^2 \beta)\times 10^{12}$ GeV, since 
the squared charged lepton Yukawa couplings in the MSSM are multiplied by this factor. Consequently, 
charged $\mu$ and $\tau$ lepton Yukawa couplings are in thermal equilibrium for
 $(1+\tan^2 \beta)\times 10^5 \: \mbox{GeV} \ll M_1 \ll (1+\tan^2 \beta)\times 10^{9} \: \mbox{GeV}$  
and all flavours in the Boltzmann equations are to be 
treated separately. For 
$(1+\tan^2 \beta)\times 10^9 \: \mbox{GeV} \ll M_1 \ll (1+\tan^2 \beta)\times 10^{12} \: \mbox{GeV}$, only the $\tau$ Yukawa coupling is in equilibrium and only the $\tau$ flavour is treated separately in the Boltzmann equations, while the $e$ and $\mu$ flavours are indistinguishable.

In what follows we will concern ourselves mainly with the regime where all flavours in the Boltzmann equations are to be treated separately, i.e.\ 
$10^5 \: \mbox{GeV} \ll M_1 \ll 10^{9} \: \mbox{GeV}$ in the SM and 
 $(1+\tan^2 \beta)\times 10^5 \: \mbox{GeV} \ll M_1 \ll (1+\tan^2 \beta)\times 10^{9} \: \mbox{GeV}$ in the MSSM. We will comment on the other two regimes and point out the differences between flavour-independent approximation and the flavour-dependent treatment, where lepton flavour is taken into account correctly in the Boltzmann equations. We start with the SM and then turn to the MSSM. 

\subsection{The Boltzmann Equations in the SM}\label{SM}
In the regime where all lepton flavours are to be treated seperately, the Boltzmann equations in the SM are given by 
\begin{eqnarray}
\label{1a}
\!\!\!\!\!\!\!\!\!\!\!\frac{\mathrm{d} Y_{N_1}}{\mathrm{d} z} \!\!&=&\!\! 
  \frac{-z}{s H(M_1)} (\gamma_D + \gamma_{\mathrm{S},\Delta L = 1})
\left(  \frac{Y_{N_1}}{Y^\mathrm{eq}_{N_1}} - 1  \right)\!,\\
\!\!\!\!\!\!\!\!\!\!\!\frac{\mathrm{d} Y_{\Delta_\alpha}}{\mathrm{d} z} \!\!&=&\!\! 
 \frac{-z}{s H(M_1)}  \!\left[
\varepsilon_{1,\alpha}   (\gamma_D + \gamma_{\mathrm{S},\Delta L = 1}) \left(  \frac{Y_{N_1}}{Y^\mathrm{eq}_{N_1}} - 1  \right) -
(\frac{\gamma^{\alpha}_D}{2} + \gamma^{\alpha}_{\mathrm{W},\Delta L = 1})\, \frac{\sum_\beta A_{\alpha\beta} Y_{\Delta_\beta}}{Y^\mathrm{eq}_\ell}
\right]\!,
\label{2a}
\end{eqnarray} 
where there is no sum over $\alpha$ in the last term on the right-side of Eq.~(\ref{2a}) and where $z = M_1/T$ with $T$ being the temperature. 
$Y_{N_1}$ is the density of the lightest right-handed neutrino $N_1$ with mass $M_1$.  
$Y_{\Delta_\alpha}$ are defined as $Y_{\Delta_\alpha}\equiv Y_B/3 - Y_{L_\alpha}$, where  
$Y_{L_\alpha}$ are the total lepton number densities for the flavours $\alpha = e,\mu,\tau$ and where $Y_B$ is the total baryon density. 
It is appropriate to solve the Boltzmann equations for $Y_{\Delta_\alpha}$ instead of for the number densities $Y_{\alpha}$ of the lepton doublets $\ell_\alpha$, since $\Delta_\alpha \equiv B/3 - L_\alpha$ is conserved by sphalerons and by the other SM interactions. 
For all number densities $Y$, normalization to the entropy density $s$ is understood. 
$Y^\mathrm{eq}_{N_1}$ and $Y^\mathrm{eq}_\ell$ stand for the corresponding equilibrium number densities.  
$\gamma_D$ is the thermally averaged total decay rate of $N_1$ and 
$\gamma_{\mathrm{S},\Delta L = 1}$ represents the rates for the $\Delta L = 1$ scattering processes in the thermal bath. Notice, in particular, that
$\gamma_{\mathrm{S},\Delta L = 1}$ contributes to the asymmetry, as was 
recently pointed out in 
\cite{Abada:2006ea}. 

The corresponding flavour-dependent rates for washout processes involving the lepton flavour $\alpha$ are $\gamma^{\alpha}_D$ (from inverse decays involving leptons $\ell_\alpha$) and $\gamma^{\alpha}_{\mathrm{W},\Delta L = 1}$. For brevity, we have not 
displayed further contributions from $\Delta L = 2$ scatterings, which 
can be neglected under conditions we will specify below. 
$\varepsilon_{1,\alpha}$ is the decay asymmetry of $N_1$ and $H (T)$ is the Hubble parameter.

The matrix $A$, which appears in the washout term, is defined by $Y_\alpha = \sum_\beta A_{\alpha\beta} \,Y_{\Delta_\beta}$. The values of its elements depend on which interactions, in addition to the weak and strong sphalerons, are in thermal equilibrium at the temperatures where leptogenesis takes place. 
Below $10^9$ GeV in the SM, $A$ is given by \cite{Abada:2006ea}
\begin{eqnarray}
A^\mathrm{SM} =  
\begin{pmatrix}
-151/179 & 20/179 & 20/179 \\
25/358 & -344/537 & 14/537 \\
25/358 & 14/537 & -344/537
\end{pmatrix}.
\end{eqnarray} 
Between $10^9$ and $10^{12}$ GeV in the SM, regarding the leptons only the interaction mediated by the $\tau$ Yukawa coupling is in equilibrium, and the lepton asymmetries and $B/3 - L_\alpha$ asymmetries in the $e$ and $\mu$ flavour can be combined to $Y_2 \equiv Y_{e + \mu}$ and $Y_{\Delta_2} \equiv Y_{\Delta_e+\Delta_\mu}$. In this temperature range, $A$ is given by \cite{Abada:2006ea}
\begin{eqnarray}
A^\mathrm{SM} =  
\begin{pmatrix}
-920/589 & 120/589  \\
30/589 & -390/589  \\ 
\end{pmatrix}.
\end{eqnarray} 
Above $10^{12}$ GeV in the SM we recover the flavour-independent treatment, where all asymmetries can be combined to $Y_{\Delta} \equiv Y_{\Delta_e + \Delta_\mu +  \Delta_\tau}$, and $A$ is given by $A^\mathrm{SM} = - 1$. 

Eqs.~(\ref{1a}) and (\ref{2a})
can be safely used in the range  of temperatures in which the lepton Yukawa
reactions for each flavour are fully in equilibrium. Indeed, for values
of $M_1$ close to $10^9$ GeV in the SM 
the reactions induced by  the muon 
Yukawa coupling are about to be in equilibrium and the quantum oscillations
of the asymmetries $Y_{e\mu}$ might not have been dumped fast enough
to be neglected. 
Eqs.~(\ref{2a}) may be generalized to include quantum oscillations,  
following Ref.~\cite{davidsonetal}. 
However, preliminary numerical simulations have shown that 
the off diagonal terms change the final diagonal lepton asymmetries 
by factors of order unity and therefore, from now on, we will safely
neglect them and restrict 
ourselves to Eqs.~(\ref{1a}) and (\ref{2a}).

One can implement the relevant parameters for connecting leptogenesis to neutrino models directly in the Boltzmann equations, following \cite{Abada:2006ea}. Let us therefore introduce, in addition to the decay asymmetries $\varepsilon_{1,\alpha}$, the parameters $K_{\alpha}$, which control the washout processes for the asymmetry in an individual lepton flavour $\alpha$ and $K$, which controls the source of RH neutrinos in the thermal bath, as
\begin{eqnarray}\label{Eq:Kaa}
K_{\alpha} \equiv \frac{\Gamma_{N_1 \ell_\alpha} +\Gamma_{N_1 \overline\ell_\alpha}}{H(M_1)}\;,\quad 
K \equiv \sum_\alpha  K_\alpha\;,\quad 
K_\alpha = K
\frac{(\lambda_{\nu}^{\dagger})_{1\alpha}(\lambda_{\nu})_{\alpha 1}}
{(\lambda_{\nu}^{\dagger}\lambda_{\nu})_{11}}\;.
\end{eqnarray}
$H(M_1)$ is the Hubble parameter at $T=M_1$, given by 
$H(M_1) \approx 1.66 \sqrt{g_*} M_1^2/M_p$ with $g^\mathrm{SM}_*=106.75$ being the effective number of degrees of freedom in the SM, $\lambda_{\nu}$ denotes the neutrino Yukawa matrix (using left-right notation) and $\Gamma_{N_1 \ell_\alpha}$ ($\Gamma_{N_1 \overline\ell_\alpha}$) is the decay rate of $N_1$ into Higgs and lepton doublet $\ell_\alpha$ (or conjugate final states, respectively).
The thermally averaged decay rate $\gamma_D$ is then given in terms of $\Gamma_{N_1 \ell_\alpha}$ by
\begin{eqnarray}
\gamma_D (z) = \sum_\alpha \gamma_D^\alpha \; , \quad 
\gamma^\alpha_D (z) = s \,Y^\mathrm{eq}_{N_1} \,\frac{K_1 (z)}{K_2 (z)}  \, (\Gamma_{N_1 \ell_\alpha}+\Gamma_{N_1 \overline\ell_\alpha}) \; ,
\end{eqnarray} 
where $K_1$ and $K_2$ are the modified Bessel functions of the second kind. This allows to replace
\begin{eqnarray}
\frac{\gamma_D (z)}{s H(M_1)} = K\,\frac{K_1 (z)}{K_2 (z)}  \,Y^\mathrm{eq}_{N_1}\; , \quad 
\frac{\gamma^\alpha_D (z)}{s H(M_1)} = K_{\alpha}\,\frac{K_1 (z)}{K_2 (z)}  \,Y^\mathrm{eq}_{N_1}\; , \;
\end{eqnarray} 
in Eqs.~(\ref{1a}) and (\ref{2a}). Defining in addition two functions $f_1$ and $f_2$ by  
\begin{eqnarray}
\gamma_D + \gamma_{\mathrm{S},\Delta L = 1} \equiv
\gamma_D f_1 \; , \quad
\frac{\gamma^{\alpha}_D}{2}  + \gamma^{\alpha}_{\mathrm{W},\Delta L = 1} \equiv
\gamma^{\alpha}_D f_2\; ,
\end{eqnarray}
 we can re-write the Boltzmann equations with correct flavour treatment in a simplified form as follows \cite{Abada:2006ea}:  
\begin{eqnarray}
\label{1}
\!\!\!\!\!\!\!\!\!\!\frac{\mathrm{d} Y_{N_1}}{\mathrm{d} z} \!\!&=&\!\! 
-\,  K\,z \,\frac{K_1 (z)}{K_2 (z)}\, f_1 (z) \,(Y_{N_1} - Y^\mathrm{eq}_{N_1}) \; ,\\
\!\!\!\!\!\!\!\!\!\!\frac{\mathrm{d} Y_{\Delta_\alpha}}{\mathrm{d} z} \!\!&=&\!\! 
-\,\varepsilon_{1,\alpha}\, K\,z \,\frac{K_1 (z)}{K_2 (z)}\,  f_1 (z) \,(Y_{N_1} - Y^\mathrm{eq}_{N_1}) +
K_{\alpha} \, z\, \frac{K_1 (z)}{K_2 (z)} \, f_2 (z) \,Y^\mathrm{eq}_{N_1}\,
 \frac{\sum_\beta A_{\alpha\beta} Y_{\Delta_\beta}}{Y^\mathrm{eq}_\ell} \!\; .
\label{2}
\end{eqnarray}
The  function $f_1(z)$ accounts for the 
presence of $\Delta L=1$ scatterings 
and $f_2(z)$ accounts for scatterings in the washout
term of the asymmetry \cite{lept,ogen}.  
In our numerical computations we only include processes mediated by neutrino and top Yukawa couplings, following Ref.~\cite{ogen}. This means, we neglect $\Delta L = 1$ scatterings involving gauge bosons \cite{Pilaftsis:2003gt,lept} and thermal corrections \cite{lept}, but we take into account corrections from renormalization group running between electroweak scale and $M_1$ \cite{barbieri,Antusch:2003kp}. We also neglect $\Delta L=2$ scatterings, which is a good approximation as long as 
$K_\alpha \gg 10 \times M_1 / (10^{14} \:\mbox{GeV})$ 
\cite{Abada:2006ea}. 
Finally, according to the usual assumptions for computing the damping rates in the 
Boltzmann equations, i.e.\ that elastic scattering rates are fast and that the phase space densities for both, fermions and scalars, can be approximated as 
$f(E_i,T)=(n_{i}/n^{\mathrm{eq}}_{i}) e^{-E_i/T}$, where 
$n_i^\mathrm{eq} = \tfrac{g_i}{2\pi} T m_i^2 K_2 (m_i/T)$ with $g_i$ being the number of degrees of freedom, we use  
\begin{eqnarray}
 Y^{\mathrm{eq}}_{\ell} \approx \frac{45 }{ \pi^4
 g_* } \; , \quad
 Y^{\mathrm{eq}}_{N_1}(z) \approx \frac{45}{ 2 \pi^4
 g_* } \,z^2\, K_2 (z)\;.
\end{eqnarray}	

The final lepton asymmetry in each flavour is governed by three sets of   
parameters, which can be computed within a neutrino model: 
$\varepsilon_{1,\alpha},K_{\alpha}$ 
and $K = \sum_\alpha K_{\alpha}$. 
$\varepsilon_{1,\alpha}$ are the decay asymmetries of the lightest right-handed neutrino $N_1$ into Higgs $H_\mathrm{u}$ and lepton doublet $\ell_\alpha$, defined as
\begin{eqnarray}\label{Eq:DefEps1a}
\varepsilon_{1,\alpha} = 
\frac{
\Gamma_{N_1 \ell_\alpha} - \Gamma_{N_1 \overline \ell_\alpha}
}{
\sum_\alpha (\Gamma_{N_1 \ell_\alpha} + \Gamma_{N_1 \overline \ell_\alpha})
}\; ,
\end{eqnarray} 
with the decay rates 
$\Gamma_{N_1 \ell_\alpha} = \Gamma (N_1\rightarrow H_\mathrm{u} \ell_\alpha)$ and
$\Gamma_{N_1 \overline \ell_\alpha} = \Gamma (N_1\rightarrow H^*_\mathrm{u} \overline \ell_\alpha)$.
SU(2)$_\mathrm{L}$-indices in the final state, not displayed explicitly, are summed over. 
In the SM, the tree-level decay rates are   
\begin{eqnarray}\label{Eq:treeSM}
\Gamma^\mathrm{SM}_{N_1 \ell_\alpha} +\Gamma^\mathrm{SM}_{N_1 \overline \ell_\alpha} = M_1\,\frac{(\lambda_{\nu}^{\dagger})_{1\alpha}(\lambda_{\nu})_{\alpha 1}}{8 \pi}\; .
\end{eqnarray} 
The decay asymmetry (arising at one-loop order) is \cite{Covi:1996wh,earlier}
\beq
\epsilon^\mathrm{SM}_{1,\alpha}=\frac{1}{8\pi}
\frac{\sum_{J=2,3}\mathrm{Im}\left[
(\lambda_{\nu}^{\dagger})_{1\alpha}[\lambda_{\nu}^{\dagger}\lambda_{\nu}]_{1J}(\lambda_{\nu}^T)_{J\alpha}
\right]}
{(\lambda_{\nu}^{\dagger}\lambda_{\nu})_{11}}\,
g^\mathrm{SM}\left(\frac{M_J^2}{M_1^2}\right) , \label{eq:epsaa}
\eeq
with the loop function $g$ in the SM given by 
\begin{eqnarray}
g^\mathrm{SM}(x) = \sqrt{x} \left[
\frac{1}{1-x} + 1 - (1+x)\ln\left(\frac{1+x}{x}\right)
\right] \stackrel{x \gg 1}{\longrightarrow} - \frac{3}{2 \sqrt{x}}\;.
\end{eqnarray}
Alternatively to $K_\alpha$ and $K$ defined in Eq.~(\ref{Eq:Kaa}), the parameters $\widetilde m_{1,\alpha}$ and 
$\widetilde m_1$ will be used in following, which we define as
\begin{eqnarray}\label{Eq:mtildeaa}\label{eq:mtildeaa}
\widetilde{m}_{1,\alpha } \equiv (\lambda_{\nu}^{\dagger})_{1 \alpha}(\lambda_{\nu})_{\alpha 1}\frac{v_{\rm u}^2}{M_1}
 \;  , \quad
\widetilde{m}_1 \equiv \sum_\alpha \widetilde{m}_{1,\alpha }\; ,
\end{eqnarray} 
with $v_\mathrm{u}=175\:\mbox{GeV}$. They are related to $K_\alpha$ and $K$ by
\begin{eqnarray}\label{Eq:mStar}
K = \frac{\widetilde{m}_1}{m^*}\, ,\;\: \mbox{or equivalently}\;\:
K_\alpha = \frac{\widetilde{m}_{1,\alpha}}{m^*},\;\: \mbox{with} \;\:
m_\mathrm{SM}^* \approx 1.08 \times 10^{-3} \ {\rm eV}
\, .
\end{eqnarray}

\subsection{The Boltzmann Equations in the MSSM}\label{MSSM}
In the MSSM, the density $Y_{\widetilde N_1}$ of right-handed sneutrinos as well as the densities $Y_{\widetilde \alpha}$ of the slepton doublets have to be included in the Boltzmann equations.
Denoting the total (particle and sparticle) $B/3 - L_\alpha$ asymmetries 
as $\hat Y_{\Delta_\alpha}$, the simplified Boltzmann equations are given by 
\begin{eqnarray}
\label{1s}
\frac{\mathrm{d} Y_{N_1}}{\mathrm{d} z} &=& 
- \, 2 K\,z \,\frac{K_1 (z)}{K_2 (z)}\, f_1 (z) \,(Y_{N_1} - Y^\mathrm{eq}_{N_1}) \; ,\\
\label{2s}\frac{\mathrm{d} Y_{\widetilde N_1}}{\mathrm{d} z} &=& 
- \, 2 K\,z \,\frac{K_1 (z)}{K_2 (z)}\, f_1 (z) \,(Y_{\widetilde N_1} - Y^\mathrm{eq}_{\widetilde N_1}) \; ,\\
\label{3s}  \frac{\mathrm{d} \hat Y_{\Delta_\alpha}}{\mathrm{d} z} &=& 
-\;(\varepsilon_{1,\alpha}+\varepsilon_{1,\widetilde \alpha}) \,K\, z \, 
\frac{K_1 (z)}{K_2 (z)}\,  f_1 (z) \,(Y_{N_1} - Y^\mathrm{eq}_{N_1})\nonumber \\
&&-\;
(\varepsilon_{\widetilde 1,\alpha}+\varepsilon_{\widetilde 1,\widetilde \alpha}) \,K\, z \, 
\frac{K_1 (z)}{K_2 (z)}\,  f_1 (z) \,(Y_{\widetilde N_1} - Y^\mathrm{eq}_{\widetilde N_1})
\nonumber \\
&&+\;
K_{\alpha} \, z\, \frac{K_1 (z)}{K_2 (z)} \, f_2 (z) \,
 \frac{\sum_\beta  A_{\alpha\beta} \hat Y_{\Delta_\beta}}{\hat Y^\mathrm{eq}_{\alpha}} \,(Y^\mathrm{eq}_{N_1}+Y^\mathrm{eq}_{\widetilde N_1})
\;  .
\end{eqnarray}
The matrix $A$ is defined via the relation
$\hat Y_{\alpha} = \sum_\beta A_{\alpha\beta} \, \hat Y_{\Delta_\alpha}$, with  
$\hat Y_\alpha \equiv Y_\alpha + Y_{\widetilde \alpha}$ being the combined densities for lepton and slepton doublets.

To obtain Eqs.~(\ref{1s}) - (\ref{3s}) we have made use of the fact that the tree level decay rates satisfy
\begin{eqnarray}\label{Eq:treeMSSM}
\Gamma_{N_1 \ell_\alpha} + \Gamma_{N_1 \overline \ell_\alpha}
\,=\,
\Gamma_{N_1 \widetilde \ell_\alpha} + \Gamma_{N_1 
\widetilde{\ell}_\alpha^*}
\,=\,
\Gamma_{\widetilde N_1^* \ell_\alpha}
\,=\,
\Gamma_{\widetilde N_1 \overline \ell_\alpha}
\,=\,
\Gamma_{\widetilde N_1 \widetilde \ell_\alpha}
\,=\,
\Gamma_{\widetilde N_1^* \widetilde{\ell}_\alpha^*}\; ,
\end{eqnarray}
with $\Gamma_{N_1 \ell_\alpha} \!+\! \Gamma_{N_1 \overline \ell_\alpha}$ given in Eq.~(\ref{Eq:treeSM}), 
leading to the identities
\begin{eqnarray}
K_{\alpha} 
=\! \frac{\Gamma_{N_1 \ell_\alpha} \!\!+\!\Gamma_{N_1 \overline\ell_\alpha}}{H(M_1)}
\!=\! \frac{\Gamma_{N_1 \widetilde \ell_\alpha} \!\!+\! \Gamma_{N_1 
\widetilde{\ell}_\alpha^*}}{H(M_1)} 
= \frac{\Gamma_{\widetilde N_1^* \ell_\alpha}}{H(M_1)}
= \frac{\Gamma_{\widetilde N_1 \overline \ell_\alpha}}{H(M_1)}
= \frac{\Gamma_{\widetilde N_1 \widetilde \ell_\alpha}}{H(M_1)}
= \frac{\Gamma_{\widetilde N_1^* \widetilde{\ell}_\alpha^*}}{H(M_1)} ,
\end{eqnarray}
with $K$, $K_{\alpha}$ (and $\widetilde m_1$, $\widetilde m_{1,\alpha}$) being defined  analogously to the SM case (c.f.\ Eqs.~(\ref{Eq:Kaa}) and (\ref{Eq:mtildeaa})).
From Eq.~(\ref{Eq:mStar}), we find 
\begin{eqnarray}
m_\mathrm{MSSM}^*\approx \sin^2(\beta) \times 1.58 \times 10^{-3} \ {\rm eV}  \; ,
\end{eqnarray}
using $g^{\mathrm{MSSM}}_*=228.75$ for computing $H(M_1)$. $m_\mathrm{MSSM}^*$ relates $K$, $K_{\alpha}$ to $\widetilde m_1$, $\widetilde m_{1,\alpha}$ in the MSSM by the analogous of Eq.~(\ref{Eq:mStar}).
In this conventions, the functions $f_1$ and $f_2$ in the Boltzmann equations are approximately unchanged (apart from an obvious modification in one of the scattering terms, which has only small effects). 
Note that $v_u = \sin (\beta) \times 175$ GeV in the MSSM. 
In Eqs.~(\ref{1s}) and (\ref{2s}), processes with particles and superpartners are combined, leading to the additional factor of $2$ on the right-side of the equations.
Ignoring supersymmetry breaking, the right-handed neutrinos and sneutrinos
 have equal mass $M_{1}$. 
With the usual 
approximation of taking Boltzmann statistics for both, fermions and scalars, 
we have used $Y^{\mathrm{eq}}_{\widetilde \ell} \approx Y^{\mathrm{eq}}_{\ell}$ to combine the washout terms for leptons and sleptons in the last term in Eq.~(\ref{3s}). Correspondingly, for the  
density $Y^\mathrm{eq}_{\widetilde N_1}$ of the right-handed sneutrinos we use   
\begin{eqnarray}
 Y^{\mathrm{eq}}_{\widetilde N_1}(z) \approx \frac{45}{ 2 \pi^4
 g_* } \,z^2\, K_2 (z)\;.
\end{eqnarray}

$\varepsilon_{1,\alpha}$, $\varepsilon_{1,\widetilde \alpha}$, $\varepsilon_{\widetilde 1,\alpha}$ and $\varepsilon_{\widetilde 1,\widetilde \alpha}$ are the decay asymmetries for the decay of  
 neutrino into Higgs and lepton,
 neutrino into Higgsino and slepton,
 sneutrino into Higgsino and lepton, and 
 sneutrino into Higgs and slepton, respectively,  
defined by 
\begin{eqnarray}\label{Eq:EpsMSSM_def}
\varepsilon_{1,\alpha} \!\!\!&=&\!\!\!
\frac{
\Gamma_{N_1 \ell_\alpha} - \Gamma_{N_1 \overline \ell_\alpha}
}{
\sum_\alpha (\Gamma_{N_1 \ell_\alpha} + \Gamma_{N_1 \overline \ell_\alpha})
}\; , \quad
\varepsilon_{1,\widetilde \alpha} =
\frac{
\Gamma_{N_1 \widetilde \ell_\alpha} - \Gamma_{N_1
\widetilde{\ell}_\alpha^*}
}{
\sum_\alpha (\Gamma_{N_1 \widetilde \ell_\alpha} + \Gamma_{N_1
\widetilde{\ell}_\alpha^*})
}\; , \nonumber  \\
\varepsilon_{\widetilde 1,\alpha} \!\!\!&=&\!\!\!
\frac{
\Gamma_{\widetilde N^*_1 \ell_\alpha} - \Gamma_{\widetilde N_1 \overline
  \ell_\alpha}
}{
\sum_\alpha (\Gamma_{\widetilde N^*_1 \ell_\alpha} +
\Gamma_{\widetilde N_1 \overline \ell_\alpha})
}\; , \quad
\varepsilon_{\widetilde 1,\widetilde \alpha} =
\frac{
\Gamma_{\widetilde N_1 \widetilde \ell_\alpha} - \Gamma_{\widetilde N^*_1
\widetilde{\ell}_\alpha^*}
}{
\sum_\alpha (\Gamma_{\widetilde N_1 \widetilde \ell_\alpha} +
\Gamma_{\widetilde N^*_1
\widetilde{\ell}_\alpha^*})
}\;.
\end{eqnarray}
In the MSSM, the four decay asymmetries are equal,
$\varepsilon^\mathrm{MSSM}_{1,\alpha} = 
\varepsilon^\mathrm{MSSM}_{1,\widetilde \alpha} = 
\varepsilon^\mathrm{MSSM}_{\widetilde 1,\alpha} = 
\varepsilon^\mathrm{MSSM}_{\widetilde 1,\widetilde \alpha}$, 
and given by \cite{Covi:1996wh,earlier}
\begin{eqnarray}
\varepsilon^\mathrm{MSSM}_{1,\alpha} = 
\frac{1}{8\pi}
\frac{\sum_{J=2,3}\mathrm{Im}\left[
(\lambda_{\nu}^{\dagger})_{1\alpha}[\lambda_{\nu}^{\dagger}\lambda_{\nu}]_{1J}(\lambda_{\nu}^T)_{J\alpha}
\right]}
{(\lambda_{\nu}^{\dagger}\lambda_{\nu})_{11}}\,
g^\mathrm{MSSM}\left(\frac{M_J^2}{M_1^2}\right) , 
\end{eqnarray}
with
\begin{eqnarray}
g^\mathrm{MSSM}(x) =\sqrt{x} \left[
\frac{2}{1-x} - \ln\left(\frac{1+x}{x}\right)\right]
\stackrel{x \gg 1}{\longrightarrow} - \frac{3}{\sqrt{x}}  \; .
\end{eqnarray}
The matrix $A$ depends on which MSSM interactions are in thermal equilibrium at the temperatures where leptogenesis takes place. 
Below $(1+\tan^2 \beta)\times 10^9$ GeV, where the Boltzmann equations are solved for the individual asymmetries $\hat Y_{\Delta_e}$, $\hat Y_{\Delta_\mu}$ and $\hat Y_{\Delta_\tau}$, $A$ is given by
\begin{eqnarray}
A^\mathrm{MSSM} =  
\begin{pmatrix}
-93/110 & 6/55 & 6/55 \\
3/40 & -19/30 & 1/30 \\
3/40 & 1/30 & -19/30
\end{pmatrix}.
\end{eqnarray} 
Between $(1+\tan^2 \beta)\times 10^9$ and $(1+\tan^2 \beta)\times 10^{12}$ GeV, where the relevant flavour-dependent asymmetries are $\hat Y_{\Delta_2} \equiv \hat Y_{\Delta_e + \Delta_\mu}$ and $\hat Y_{\Delta_\tau}$, we find 
\begin{eqnarray}
A^\mathrm{MSSM} =  
\begin{pmatrix}
-541/761 & 152/761  \\
46/761   & -494/761  \\ 
\end{pmatrix},
\end{eqnarray}    
and above $(1+\tan^2 \beta)\times 10^{12}$ GeV, we recover the flavour-independent treatment with $A^\mathrm{MSSM} = - 1$.

\subsection{Solving the Boltzmann Equations in the SM and MSSM}
\begin{figure}[t]
 \centering
 \includegraphics[scale=1,angle=0]{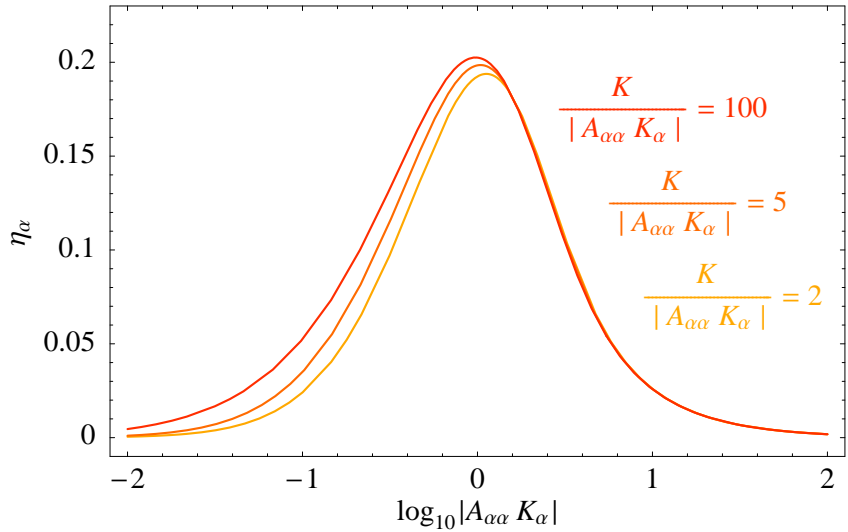}
 \caption{\label{fig:eta}
Flavour-dependent efficiency factor $\eta(A_{\alpha\alpha}K_{\alpha},K)$ in the MSSM as a function of $A_{\alpha\alpha}K_{\alpha}$, for fixed values of $K/|A_{\alpha\alpha}K_{\alpha}| = 2,5$ and $100$, obtained from solving the flavour-dependent Boltzmann equations \cite{Abada:2006ea} generalized to the MSSM (with $\tan \beta = 50$, as an example), as displayed in Eqs.~(\ref{1s}) - (\ref{3s}). 
For larger $K/|A_{\alpha\alpha}K_{\alpha}|$ the plot looks virtually like for $K/|A_{\alpha\alpha}K_{\alpha}| = 100$. More relevant than the differences in the flavour-dependent efficiency factors for different $K/|A_{\alpha\alpha}K_{\alpha}|$ is that the total baryon asymmetry is the sum of each individual lepton asymmetries, which is weighted by the corresponding efficiency factor. As explained in the text, this can change the amount of the produced baryon asymmetry dramatically, compared to the flavour-independent approximation 
\cite{davidsonetal,nardietal,Abada:2006ea}. }
\end{figure}

Solving the Boltzmann equations, for $z$ from $0$ to $\infty$, in the
SM or in the MSSM yields the final $B/3 - L_\alpha$ asymmetries $Y^\mathrm{SM}_{\Delta_\alpha}$ or
$\hat Y^\mathrm{MSSM}_{\Delta_\alpha}$  
in the individual flavours. 
It is convenient to parameterize the produced asymmetries in terms of 
an efficiency factor $\eta_{\alpha}$ which, in the approximation that the small off-diagonal elements of $A$ are neglected, is a function
of $A_{\alpha\alpha}K_{\alpha}$ (no sum over $\alpha$) and $K$, i.e.\ $\eta_{\alpha} = \eta (A_{\alpha\alpha}K_{\alpha},K)$, as
\begin{eqnarray}\label{Eq:eta_aa}
Y^\mathrm{SM}_{\Delta_\alpha} &=& \eta^\mathrm{SM}_{\alpha} \,\varepsilon^\mathrm{SM}_{1,\alpha} \,
Y^{\mathrm{eq}}_{N_1}(z\ll 1)\;, \\
\label{Eq:eta_aa_MSSM} \hat Y^\mathrm{MSSM}_{\Delta_\alpha} &=& \eta^\mathrm{MSSM}_{\alpha} \,
\varepsilon^\mathrm{MSSM}_{1,\alpha}\, \left[ 
Y^{\mathrm{eq}}_{N_1}(z\ll 1) +Y^{\mathrm{eq}}_{\widetilde N_1}(z\ll 1)
\right] ,\label{Eq:eta_aa_MSSM}
\end{eqnarray}
generalizing the notation of \cite{lept} to the flavour-dependent treatment. 
Beyond the approximations of Secs.~\ref{SM} and \ref{MSSM}, the efficiency factors also depend on $M_1$, and they depend on $\tan \beta$ in the MSSM.
  $Y^{\mathrm{eq}}_{N_1}(z\ll 1)$ and 
 $Y^{\mathrm{eq}}_{\widetilde N_1}(z\ll 1)$ are the 
 number densities of the neutrino and sneutrino at $T \gg M_{1}$, 
 if they were in thermal 
 equilibrium, normalized with respect to the entropy
 density. In the Boltzmann approximation, they are given by
\begin{eqnarray}
 Y^{\mathrm{eq}}_{N_1}(z\ll 1) \;\approx \;Y^{\mathrm{eq}}_{\widetilde N_1}(z\ll 1) \;\approx\; \frac{45}{ \pi^4
 g_* }.
 \end{eqnarray}
Eqs.~(\ref{Eq:eta_aa}) and (\ref{Eq:eta_aa_MSSM}) define the flavour-dependent efficiency factor in the SM and in the MSSM, and it can be computed by means of the Boltzmann equations \cite{davidsonetal,Abada:2006ea} and its MSSM generalizations in Eqs.~(\ref{1s}) - (\ref{3s}), where lepton flavour is taken into account correctly. 
The equilibrium number densities in Eqs.~(\ref{Eq:eta_aa}) and (\ref{Eq:eta_aa_MSSM}) serve as a normalization.  
A thermal population $N_1$ (and $\widetilde N_1$) decaying completely out of equilibrium (without washout effects) would lead to $\eta_\alpha = 1$.
Of course, $K/K_{\alpha} \ge 1$ always holds by definition 
but $K$ can be significantly larger than $K_{\alpha}$. $\eta(A_{\alpha\alpha}K_{\alpha},K)$ is shown as a function of $A_{\alpha\alpha}K_\alpha$ in the MSSM for fixed values of $K/|A_{\alpha\alpha}K_{\alpha}| = 2,5$ and $100$ in Fig.~\ref{fig:eta}. In the SM, $\eta$ has the same qualitative (and a similar quantitative) behavior.

What is more relevant than the differences in the flavour-dependent efficiency factors (c.f.\ Fig.~\ref{fig:eta}) is that the total baryon asymmetry is the sum of each individual lepton asymmetries, which is weighted by the corresponding efficiency factor \cite{davidsonetal,nardietal,Abada:2006ea}. Therefore,
upon summing over the lepton asymmetries, the total baryon number is
generically not proportional to the sum over the
CP asymmetries, 
$\varepsilon_1 = \sum_\alpha \varepsilon_{1,\alpha}$
as in the flavour-independent 
approximation where the lepton flavour is 
neglected in the Boltzmann equations. In Eq.~(\ref{2}), this corresponds 
to replacing $Y_{\Delta_\alpha}$ by the ``one-single'' flavour 
$Y_\Delta= Y_{\Delta_e + \Delta_\mu + \Delta_\tau}$ (the total lepton asymmetry), 
the flavour 
dependent decay asymmetries by $\varepsilon_1$, the washout 
parameters by 
$K = \sum_\alpha K_{\alpha}$, and the matrix $A$ by $-1$. 
The produced asymmetry in the flavour-independent approximation is then given by
$Y = \varepsilon_1 \,\eta^\mathrm{ind} (K)\,Y^\mathrm{eq}_{N_1}(z\ll 1)$, 
where the flavour-independent efficiency factor $\eta^\mathrm{ind} (K)$ is related to 
$\eta (A_{\alpha\alpha}K_{\alpha},K)$ in Eq.~(\ref{Eq:eta_aa}) by 
$\eta^\mathrm{ind} (x) = \eta(-x,x)$. 
In other words, in the flavour-independent approximation the total baryon asymmetry is a function of
$\left(\sum_\alpha \varepsilon_{1,\alpha}\right)\times \eta^\mathrm{ind}\ (\sum_\beta 
K_{\beta})$. In the correct flavour treatment
the baryon asymmetry is a function of $\sum_\alpha 
\varepsilon_{1,\alpha}\eta\left(A_{\alpha\alpha}K_{\alpha},K\right)$.
If $N_1$ decays about equally to all flavours and produces
about the equal asymmetry in all flavours, then (neglecting the effects of $A$ for the moment) one expects that the 
flavour-independent approximation underestimates the asymmetry by roughly a 
factor of three for the case of 
strong washout and overestimates by a factor of three for the case of weak wash out. 
The approximate factor of three can be understood with the analytic approximations presented in \cite{Abada:2006ea}, 
from which we can see that in the case of weak (strong) washout for all flavours, the efficiencies are roughly (inverse) proportional to $K_{\alpha}$.
However, as we will see, there are situations in the SD 
models where the difference is much more dramatic.

Let us note at this point how to generalize the above discussion for the range 
$10^9 \: \mbox{GeV} \ll M_1 \ll 10^{12} \: \mbox{GeV}$ in the SM and  
$(1+\tan^2 \beta)\times 10^9 \: \mbox{GeV} \ll M_1 \ll (1+\tan^2 \beta)\times 10^{12} \: \mbox{GeV}$ in the MSSM, where only the $\tau$ is in equilibrium and is treated separately in the Boltzmann equations, while the $e$ and $\mu$ flavours are indistinguishable. In this case, following \cite{davidsonetal,Abada:2006ea}, one can combine the asymmetries for the $e$ and $\mu$ flavours to a combined density $Y_{\Delta_2} = Y_{\Delta_e + \Delta_\mu}$ in the Boltzmann equations, where $K_\alpha$ is substituted by $K_2 = K_e + K_\mu$. The corresponding decay asymmetry is $\varepsilon_{1,2} = \varepsilon_{1,e} + \varepsilon_{1,\mu}$. 
Above $10^{12} \: \mbox{GeV}$ in the SM and above $(1+\tan^2 \beta)\times 10^{12} \: \mbox{GeV}$ in the MSSM, we can combine all asymmetries for the $e,\mu$ and $\tau$  flavours to a combined density $Y_\Delta = Y_{\Delta_e + \Delta_\mu + \Delta_\tau}$ in the Boltzmann equations and substitute $K_\alpha$ by $K = K_e + K_\mu + K_\tau$. The decay asymmetry then reduces to the flavour-independent one, $\varepsilon_{1} = \varepsilon_{1,e} + \varepsilon_{1,\mu} + \varepsilon_{1,\tau}$.

The produced lepton asymmetries are partly converted into a final baryon asymmetry by 
sphalerons. For all temperature ranges, the produced baryon asymmetry (normalized to the entropy density) can be computed from the densities $Y^\mathrm{SM}_{\Delta_\alpha}$ and $\hat Y^\mathrm{MSSM}_{\Delta_\alpha}$ in the SM and MSSM, respectively, as
\begin{eqnarray}\label{Eq:YB3f}
Y^\mathrm{SM}_B &=& \frac{12}{37} \sum_\alpha Y^\mathrm{SM}_{\Delta_\alpha} \; ,\\
Y^\mathrm{MSSM}_B &=& \frac{10}{31} \sum_\alpha \hat Y^\mathrm{MSSM}_{\Delta_\alpha}\;. 
\end{eqnarray}

\section{Sequential Dominance}\label{SD}

To understand how sequential dominance works, we begin by
writing the right-handed neutrino Majorana mass matrix $M_{\mathrm{RR}}$ in
a diagonal basis as
\begin{equation}
M_{\mathrm{RR}}=
\begin{pmatrix}
M_A & 0 & 0 \\
0 & M_B & 0 \\
0 & 0 & M_C%
\end{pmatrix}.
\end{equation}
Note that, as stated earlier, we work in a basis where the charged lepton Yukawa matrix 
is diagonal. 
In this basis we write the neutrino (Dirac) Yukawa matrix $\lambda_{\nu}$ in
terms of $(1,3)$ column vectors $A_i,$ $B_i,$ $C_i$ as
\begin{equation}
\lambda_{\nu }=
\begin{pmatrix}
A & B & C
\end{pmatrix},
  \label{Yukawa}
\end{equation}
using left-right convention as in Eqs.~(\ref{Eq:L}) and (\ref{Eq:W}). 
The Dirac neutrino mass matrix is then given by $m_{\mathrm{LR}}^{\nu}=\lambda_{\nu}v_{\mathrm{
u}}$. The term for the light neutrino masses in the effective Lagrangian (after electroweak symmetry breking), resulting from integrating out the massive right
handed neutrinos, is
\begin{equation}
\mathcal{L}^\nu_{eff} = \frac{(\nu_{i}^{T} A_{i})(A^{T}_{j} \nu_{j})}{M_A}+\frac{(\nu_{i}^{T} B_{i})(B^{T}_{j} \nu_{j})}{M_B}
+\frac{(\nu_{i}^{T} C_{i})(C^{T}_{j} \nu_{j})}{M_C}  \label{leff}
\end{equation}
where $\nu _{i}$ ($i=1,2,3$) are the left-handed neutrino fields.
Sequential dominance then corresponds to the third
term being negligible, the second term subdominant and the first term
dominant:
\beq\label{SDcond}
\frac{A_{i}A_{j}}{M_A} \gg
\frac{B_{i}B_{j}}{M_B} \gg
\frac{C_{i}C_{j}}{M_C} \, .
\label{SD1}
\eeq
In addition, we shall shortly see that small $\theta_{13}$ 
and almost maximal $\theta_{23}$ require that 
\beq
|A_1|\ll |A_2|\approx |A_2|.
\label{SD2}
\eeq
Without loss of generality, then, we shall label the dominant
right-handed neutrino and Yukawa couplings as $A$, the subdominant
ones as $B$, and the almost decoupled (subsubdominant) ones as $C$. 
Note that the mass ordering of right-handed neutrinos is 
not yet specified. Again without loss of generality we shall 
order the right-handed neutrino masses as $M_1<M_2<M_3$,
and subsequently identify $M_A,M_B,M_C$ with $M_1,M_2,M_3$
in all possible ways. The following results for the masses and mixing
angles are independent of the mass ordering of the right-handed neutrinos.

Writing 
$A_\alpha = |A_\alpha| e^{i \phi_{A_1}}$, 
$B_\alpha = |B_\alpha| e^{i \phi_{B_1}}$, 
$C_\alpha = |C_\alpha| e^{i \phi_{C_1}}$ and working in the mass basis
of the charged leptons, under the SD condition Eq.~(\ref{SD1}),
we obtain for the lepton mixing angles 
\cite{King:1998jw}:
\begin{subequations}\label{anglesSD}\begin{eqnarray}
\label{Eq:t23} 
\tan \theta_{23} &\approx& \frac{|A_2|}{|A_3|}\;, \\
\label{Eq:t12}
\tan \theta_{12} &\approx& 
\frac{|B_1|}{c_{23}|B_2|\cos \tilde{\phi}_2 - 
s_{23}|B_3|\cos \tilde{\phi}_3  } \;,\\
\label{Eq:t13}
\theta_{13} &\approx& 
e^{i (\tilde{\phi} + \phi_{B_1} - \phi_{A_2})}
\frac{|B_1| (A_2^*B_2 + A_3^*B_3) }{\left[|A_2|^2 + |A_3|^2\right]^{3/2} }
\frac{M_A}{M_B} 
+\frac{e^{i (\tilde{\phi} + \phi_{A_1} - \phi_{A_2})} |A_1|}
{\sqrt{|A_2|^2 + |A_3|^2}} ,
\end{eqnarray}\end{subequations}
and for the masses
\begin{subequations}\label{massesSD}\begin{eqnarray}
\label{Eq:m3} m_3 &\approx& \frac{(|A_2|^2 + |A_3|^2)v^2}{M_A}\;, \\
\label{Eq:m2} m_2 &\approx& \frac{|B_1|^2 v^2}{s^2_{12} M_B}\;, \\
\label{Eq:m1}m_1 &\approx& {\cal O}(|C|^2 v^2/M_C) \;. 
\end{eqnarray}\end{subequations}
We would like to note that tri-bimaximal mixing \cite{tribi} corresponds to the choice
\cite{King:2005bj}:
\begin{eqnarray}
|A_{1}| &=&0,  \label{tribicondsd} \\
\text{\ }|A_{2}| &=&|A_{3}|,  \label{tribicondse} \\
|B_{1}| &=&|B_{2}|=|B_{3}|,  \label{tribicondsa} \\
A^{\dagger }B &=&0.  \label{zero}
\end{eqnarray} 
This is called constrained sequential dominance (CSD) \cite{King:2005bj}.

Let us now turn to the issue of leptonic CP phases in SD. In
particular we will consider the MNS phase $\delta$,
relevant for neutrino oscillations,
and then study in Sec.~\ref{Sec:MNSLGlink} how it is linked to CP violation required for leptogenesis. 
As in \cite{King:2002qh}
the MNS phase $\delta$ is fixed by the requirement that we have already 
imposed in Eq.~(\ref{Eq:t12}) that $\tan(\theta_{12})$ is real and positive,
\begin{eqnarray}
\lefteqn{ \label{real12} c_{23}|B_2|\sin \tilde{\phi}_2 \;\approx\; s_{23}|B_3|\sin \tilde{\phi}_3 \; ,}  \\
\lefteqn{\label{pos12} c_{23}|B_2|\cos \tilde{\phi}_2 - s_{23}|B_3|\cos \tilde{\phi}_3 \;>\; 0\; ,}
\end{eqnarray} 
where
\bea
\tilde{\phi}_2 & \equiv & \phi_{B_2}-\phi_{B_1}-\tilde{\phi}+\delta\; ,
\nonumber \\
\tilde{\phi}_3 & \equiv & \phi_{B_3}-\phi_{B_1}+\phi_{A_2}-\phi_{A_3}
-\tilde{\phi}+\delta \; .
\label{tildephi23}
\eea
The phase $\tilde{\phi}$ is fixed by the requirement (not yet imposed
in Eq.~(\ref{Eq:t13}))
that the angle $\theta_{13}$ is real and positive.
In general this condition is rather complicated since the expression
for $\theta_{13}$ is a sum of two terms.
However if, for example, $A_1=0$ then $\tilde{\phi}$ is fixed by:
\beq
\tilde{\phi}\approx \phi_{A_2}-\phi_{B_1}-\zeta
\label{tildephi}
\eeq
where 
\beq
\zeta = \arg\left(A_2^*B_2 + A_3^*B_3  \right).
\label{eta}
\eeq
Eq.~(\ref{eta}) may be expressed as 
\beq
\tan \zeta \approx \frac{|B_2|s_{23}s_2+|B_3|c_{23}s_3}
{|B_2|s_{23}c_2+|B_3|c_{23}c_3}\,.
\label{taneta}
\eeq
Inserting $\tilde{\phi}$ of Eq.~(\ref{tildephi}) into 
Eqs.~(\ref{real12}), (\ref{tildephi23}), we obtain a relation
which can be expressed as
\beq
\tan (\zeta +\delta) \approx \frac{|B_2|c_{23}s_2-|B_3|s_{23}s_3}
{-|B_2|c_{23}c_2+|B_3|s_{23}c_3}\,.
\label{tanetadelta}
\eeq
In Eqs.~(\ref{taneta}), (\ref{tanetadelta}) we have written
$s_i=\sin \zeta_i$, $c_i=\cos \zeta_i$, where we have defined
\beq
\zeta_2\equiv \phi_{B_2}-\phi_{A_2}\;, \ \ \zeta_3\equiv
\phi_{B_3}-\phi_{A_3}\;,
\label{eta23}
\eeq
which are invariant under a charged lepton phase transformation.
The reason why the seesaw parameters only involve two invariant
phases rather than the usual six, is due to the SD assumption
in Eq.~(\ref{SD1})
that has the effect of effectively decoupling the right-handed neutrino 
of mass $M_C$ from the seesaw mechanism, which removes three phases,
together with the further assumption (in this case) of
$A_1=0$, which removes another phase.

\section{Flavour Matters in Leptogenesis with Sequential Dominance}\label{Sec:FlavourLGinSD}

Let us now consider leptogenesis in neutrino mass models with sequential dominance (SD), taking into account lepton flavour in the Boltzmann equations. In SD, there are three classes of models with different characteristic predictions for leptogenesis. They differ by the role of the lightest right-handed neutrino in SD, which can either be the dominant one $M_1 = M_A$, the subdominant one $M_1 = M_B$ or the subsubdominant one $M_1 = M_C$ (which, in the SD limit, only contributes to $m_1$ but has a negligible effect on the neutrino mixing angles and CP phases). 
The possible form of the neutrino Yukawa matrix $\lambda_\nu$ is then given by
\begin{eqnarray}
\lambda_\nu &=& (A,B,C)\; \mbox{or} \;\lambda_\nu = (A,C,B)\;,\;\, \mbox{for $M_1 = M_A$,}\\
\lambda_\nu &=& (B,A,C)\; \mbox{or} \;\lambda_\nu = (B,C,A)\;,\;\, \mbox{for $M_1 = M_B$,}\\
\lambda_\nu &=& (C,A,B)\; \mbox{or} \;\lambda_\nu = (C,B,A)\;,\;\, \mbox{for $M_1 = M_C$,}
\end{eqnarray}
using the notation of Sec.~\ref{SD}, where we have ordered the columns
according to $M_{RR}=\mbox{diag}(M_1,M_2,M_3)$ where $M_1<M_2<M_3$. 
The flavour specific decay asymmetries $\varepsilon_{1,\alpha}$ and washout parameters $\widetilde m_{1,\alpha}$,  calculated from Eqs.~(\ref{eq:epsaa}) and (\ref{eq:mtildeaa}), 
are given in Tab.~\ref{tab:DecayAsym} for the three classes of SD.   
  
\begin{table} 
  \centering 
  \begin{eqnarray*} 
  \begin{array}{|c||c|c|} 
 \hline 
\mbox{Type of SD}
&
\varepsilon_{1,\alpha}   \vphantom{\frac{f^f}{f^f}}
&
\widetilde m_{1,\alpha}
\\
\hline\hline
M_1 = M_A
&
\displaystyle -\frac{3 M_1}{16 \pi } \left\{ \frac{\mathrm{Im}\left[ A_\alpha^* B_\alpha  (A^\dagger B) \right]}{ M_B\,(A^\dagger A)}
+
\frac{\mathrm{Im}\left[ A_\alpha^* C_\alpha  (A^\dagger C) \right]}{ M_C\,(A^\dagger A)}
\right\}\! 
                        \vphantom{\frac{\frac{f^f}{f}}{\frac{f}{f}}}
&
\displaystyle \frac{|A_\alpha|^2 v_\mathrm{u}^2}{M_1}
\\
\hline
M_1 = M_B
&
\displaystyle  -\frac{3 M_1}{16 \pi } \left\{\frac{\mathrm{Im}\left[ B_\alpha^* A_\alpha  (B^\dagger A) \right]}{ M_A\,(B^\dagger B)}
+
\frac{\mathrm{Im}\left[ B_\alpha^* C_\alpha  (B^\dagger C) \right]}{ M_C\,(B^\dagger B)}
\right\}\!   
                       \vphantom{\frac{\frac{f^f}{f}}{\frac{f}{f}}}
& 
\displaystyle \frac{|B_\alpha|^2 v_\mathrm{u}^2}{M_1}
\\
\hline 
M_1 = M_C
&
\displaystyle  -\frac{3 M_1}{16 \pi } \left\{
\frac{\mathrm{Im}\left[ C_\alpha^* A_\alpha (C^\dagger A) \right]}{ M_A (C^\dagger C)} 
+
\frac{\mathrm{Im}\left[ C_\alpha^* B_\alpha (C^\dagger B) \right]}{ M_B (C^\dagger C)}
\right\}\! 
                       \vphantom{\frac{\frac{f^f}{f}}{\frac{f}{f_f}}}
&
\displaystyle \frac{|C_\alpha|^2 v_\mathrm{u}^2}{M_1}
\\
\hline 
\end{array} 
\end{eqnarray*} 
\caption{
\label{tab:DecayAsym} Flavour specific decay asymmetries $\varepsilon_{1,\alpha}$ for the decay of the lightest right-handed neutrino with mass $M_1$ (in the limit $M_1 \ll M_2 , M_3$) and washout parameters $\widetilde m_{1,\alpha}$ in classes of
models with sequential right-handed neutrino dominance (SD) in the SM. The MSSM decay asymmetries are a factor of $2$ larger, in the considered limit.} 
\end{table}

\subsection{Decay Asymmetries and Washout Parameters}
\label{Sec:DecayAndWashout}
We can now derive theoretical expectations and constraints on the
flavour specific decay asymmetries and washout parameters from
requiring consistency with the present neutrino data in the classes of
neutrino mass models discussed above. The experimental results from
neutrino oscillations indicate nearly maximal mixing $\theta_{23}
\approx 45^\circ$, large, but non-maximal mixing $\theta_{12} \approx
33^\circ$ and small mixing $\theta_{13} \lesssim 13^\circ$, as well as
small neutrino mass squared differences $\Delta m^2_{31} = m_3^2 -
m_1^2 \approx 2.2 \times 10^{-3}$ eV$^2$ and $\Delta m^2_{21} = m_2^2 -
m_1^2 \approx 8.1 \times 10^{-5}$ eV$^2$ \cite{Maltoni:2004ei}. In
neutrino mass models satisfying SD, the neutrino Yukawa couplings and
the masses of the right-handed neutrinos are related to the neutrino
masses and mixings by Eqs.~(\ref{anglesSD}) and (\ref{massesSD}).

We will focus first on the case where all three flavours are treated separately in the Boltzmann equations. As discussed in Sec.~\ref{Sec:FML}, this is the case for  
temperatures below $10^{9} \: \mbox{GeV}$ in the SM and below $(1+\tan^2 \beta)\times 10^{9} \: \mbox{GeV}$ in the MSSM. Typical examples where this case is relevant are thus   unified models of flavour with large $\tan \beta$. 
We will then comment on the two remaining cases, where (i) the $e$ and $\mu$ flavour are combined in the Boltzmann equations and only the $\tau$ is treated independently and (ii) where all flavours are combined and the treatment becomes flavour-independent, in Sec.~\ref{Sec:2otherCases}. 

\subsubsection{Properties of the Decay Asymmetries $\varepsilon_{1,\alpha}$}\label{Sec:Decay}

The decay asymmetries $\varepsilon_{1,\alpha}$ of
Tab.~\ref{tab:DecayAsym} contain contributions from the two heavier
right-handed neutrinos with masses $M_2$ and $M_3$. Using the SD
conditions of Eq.~(\ref{SDcond}), we see that the largest contributions to
the decay asymmetries stem from the terms in Tab.~\ref{tab:DecayAsym}
containing $M_A$, followed by the contributions containing $M_B$, and
finally terms containing $M_C$ give the smallest contributions which
vanish in the limit $m_1 \to 0$. In the following, we will neglect the
contributions to the decay asymmetry containing $M_C$, keeping in mind
that they may be important when the leading contributions are
suppressed, for instance due to a configuration of complex phases or
due to a particular structure of the neutrino Yukawa couplings $A_i$
and $B_i$.  In the following, we will compare the flavour specific
decay asymmetries with the maximally possible decay asymmetry
\begin{eqnarray}
\varepsilon^{\mathrm{max}} = \frac{3}{16 \pi} \frac{M_1}{v_\mathrm{u}^2} m_3 \; ,
\end{eqnarray} 
which provides an upper bound on the decay asymmetry $\varepsilon_1$ in the SM in the flavour-independent approximation. In the MSSM, the maximal decay asymmetry is a factor of 2 larger. The flavour specific decay asymmetries can be constrained in the three cases as follows: 

\begin{itemize}
\item In the case $M_1 = M_A$, assuming that the 
sub-dominant right-handed neutrino provides the leading contribution to the decay asymmetry and that the term containing $M_C$ can be neglected, the flavour-specific  asymmetries are given by
\begin{eqnarray}
\varepsilon_{1,\alpha} 
\approx -\frac{3 M_1}{16 \pi } 
\frac{\mathrm{Im}\left[ A_\alpha^* B_\alpha  (A^\dagger B) \right]}{ M_B\,(A^\dagger A)}\;.
\end{eqnarray}
Using $|A_1| \ll |A_2| \approx |A_3|$, as suggested by and Eqs.~(\ref{Eq:t13}) and
(\ref{Eq:t23}) and the smallness of $\theta_{13}$ together with the approximate maximality of $\theta_{23}$, we can constrain $\varepsilon_{1,\beta}$ with $\beta\in\{\mu,\tau\}$ and $\varepsilon_{1,e}$ as follows: 
\begin{eqnarray}
|\varepsilon_{1,\beta}| 
&\approx& \left|\frac{3 M_1}{16 \pi } 
\frac{\mathrm{Im}\left[ A_\beta^* B_\beta  (A^\dagger B) \right]}{ M_B\,(A^\dagger A)}\right|
\;\le\;
\left|\frac{3 M_1}{16 \pi } 
\frac{\left[\, |A_3| |B_\beta| |A_3|  (|B_2| + |B_3|) \right]}{ M_B\,2 |A_3|^2}\right|\nonumber \\
&\le& 
\left|\frac{3 M_1}{16 \pi } \frac{1}{2}
\frac{\left[ \, |B_\beta|   (|B_2| + |B_3|) \right]}{ M_B} \right|
\;\approx\; \varepsilon^{\mathrm{max}} \label{eq:MAepsbb} \, 
{\cal O}\!\left(\frac{m_2}{m_3}\right), \\
|\varepsilon_{1,e}| 
&\lesssim& \varepsilon^{\mathrm{max}}  \, {\cal O}\! \left(\frac{m_2}{m_3} \right)\, \frac{|A_1|}{|A_3|} \left|\frac{\left[ \, |B_1|   (|B_2| + |B_3|) \right]}{ M_B} \right|
\;\ll\;   \varepsilon^{\mathrm{max}} \, {\cal O}\! \left(\frac{m_2}{m_3} \right)\label{eq:MAepsee} .
\end{eqnarray}
The estimate in the last step of Eqs.~(\ref{eq:MAepsbb}) and
(\ref{eq:MAepsee}) is based on the SD conditions in
Eq.~(\ref{SD1}) and on Eq.~(\ref{Eq:t12}). The latter states
that in order to generate large mixing $\theta_{12} \approx 33^\circ$,
$|B_2|$ and/or $|B_3|$ have to be of the same order as $|B_1|$, which
in turn is related to $m_2$ by Eq.~(\ref{Eq:m2}).  We conclude that
$|\varepsilon_{1,\mu}|$ and $|\varepsilon_{1,\tau}|$ are somewhat
suppressed compared to the maximally possible value
$\varepsilon^{\mathrm{max}}$ as was found for $\varepsilon_1 = \sum_\alpha \varepsilon_{1,\alpha}$ in the flavour-independent treatment
in \cite{Hirsch:2001dg}.  The value of $|\varepsilon_{1,e}|$
given in Eq.~(\ref{eq:MAepsee}) is suppressed compared to
$|\varepsilon_{1,\mu}|$ and $|\varepsilon_{1,\tau}|$ due to $|A_1|
\ll |A_2|,|A_3|$ and it is related to the mixing angle $\theta_{13}$,
as can be seen from Eq.\ (\ref{Eq:t13}).

\item In the case $M_1 = M_B$, using again $|A_1| \ll| A_2| \approx |A_3|$, we find that we can constrain the decay asymmetries 
$\varepsilon_{1,\beta}$, with $\beta \in \{\mu,\tau\}$, and $\varepsilon_{1,e}$ as
\begin{eqnarray}
|\varepsilon_{1,\beta}| 
&\approx& \left| 
\frac{3 M_1}{16 \pi } 
\frac{\mathrm{Im}\left[ B_\beta^* A_\beta  (B^\dagger A) \right]}{ M_A\,(B^\dagger B)}
\right|
\;\le\;
\left|\frac{3 M_1}{16 \pi } 
\frac{\left[ \,|B_\beta| \,|A_3|^2  (|B_2| + |B_3|) \right]}{M_A\,(B^\dagger B)}\right|\nonumber \\
&=& 
 \varepsilon^{\mathrm{max}}\,\frac{1}{2} \,
\left|\frac{\left[ \, |B_\beta|   (|B_2| + |B_3|) \right]}{(B^\dagger B)} \right|
\;\le\; \varepsilon^{\mathrm{max}}  \label{eq:MBepsbb}\\
|\varepsilon_{1,e}| 
&\lesssim&  \varepsilon^{\mathrm{max}}  \,\frac{1}{2}\, \frac{|A_1|}{|A_3|} 
\left|\frac{\left[ \, |B_1|   (|B_2| + |B_3|) \right]}{(B^\dagger B)} \right|
\;\ll\; \varepsilon^{\mathrm{max}} \label{eq:MBepsee}\;.
\end{eqnarray}
For $|\varepsilon_{1,\beta}|$, the maximal value $\varepsilon^{\mathrm{max}}$ can be nearly saturated, whereas $|\varepsilon_{1,e}|$ is suppressed due to $|A_1| \ll| A_2| ,|A_3|$ as in the case $M_1=M_A$.

\item In the case $M_1 = M_C$, the decay asymmetries
$\varepsilon_{1,\alpha}$ obtain contributions from the right-handed
neutrinos with masses $M_A$ and $M_B$.  Using analogous estimates than
for the case $M_1 = M_B$, we can show that the contribution from the
right-handed neutrino with mass $M_A$ allows decay asymmetries
$|\varepsilon_{1,\mu}|$ and $|\varepsilon_{1,\tau}|$ close to the
upper bound $\varepsilon^{\mathrm{max}}$ similar to the result for $\varepsilon_1$ in the
flavour-independent treatment in \cite{Hirsch:2001dg},
but due to $|A_1| \ll |A_2|
\approx |A_3|$ the decay asymmetry $\varepsilon_{1,e}$ is suppressed.
From $M_B$, using estimates as for the case $M_1 = M_A$, all decay
asymmetries can get contributions $|\varepsilon_{1,\alpha}| \lesssim
{\cal O}(m_2/m_3) \,\varepsilon^{\mathrm{max}}$. We will return to the
case $M_1 = M_C$ in Sec.~\ref{Sec:CSDinMC}, where we discuss the
specific example of tri-bimaximal mixing via constrained sequential
dominance (CSD). There we will also present explicit formulae for the
decay asymmetries which illustrate the above statements.
\end{itemize}

\noindent The above derived constraints on the decay asymmetries $\varepsilon_{1,\alpha}$ in the classes of models under consideration are summarized in Tab.~\ref{tab:PropertiesEpsSD}.

\begin{table} 
  \centering 
  \begin{eqnarray*} 
  \begin{array}{|c||c|c|c|} 
 \hline 
\mbox{Type of SD}
&
\mbox{relation for}\;  \varepsilon_{1,e}  \vphantom{\frac{f^f}{f^f}}
&
\mbox{relation for}\;  \varepsilon_{1,\mu}
&
\mbox{relation for}\;  \varepsilon_{1,\tau}
\\
\hline\hline
M_1 = M_A  \vphantom{\frac{f^f}{f^f}}
&
\ll      {\cal O}(\tfrac{m_2}{m_3}) \varepsilon^{\mathrm{max}}
& 
\lesssim {\cal O}(\tfrac{m_2}{m_3}) \varepsilon^{\mathrm{max}}
&
\lesssim {\cal O}(\tfrac{m_2}{m_3}) \varepsilon^{\mathrm{max}}
\\
\hline 
M_1 = M_B  \vphantom{\frac{f^f}{f^f}}
&
\ll  \varepsilon^{\mathrm{max}}
&
\lesssim \varepsilon^{\mathrm{max}}
&
\lesssim \varepsilon^{\mathrm{max}}
\\
\hline
M_1 = M_C  \vphantom{\frac{f^f}{f^f}}
&
\lesssim {\cal O}(\tfrac{m_2}{m_3}) \varepsilon^{\mathrm{max}}
&
\lesssim  \varepsilon^{\mathrm{max}}
&
\lesssim  \varepsilon^{\mathrm{max}}
\\
\hline 
\end{array} 
\end{eqnarray*} 
\caption{
\label{tab:PropertiesEpsSD} Constraints and theoretical expectations on the flavour specific decay asymmetries from consistency with low energy neutrino observables in classes of models with SD in the SM, in the limit $M_1 \ll M_2 , M_3$. 
$\varepsilon^{\mathrm{max}}$, the upper bound on the decay asymmetry, is given by $\frac{3}{16 \pi} \frac{M_1}{v_\mathrm{u}^2} m_3$. In the MSSM, the decay asymmetries are a factor of $2$ larger.} 
\end{table}

\subsubsection{Properties of the Washout Parameters $\widetilde m_{1,\alpha}$}\label{Sec:Washout}

\begin{itemize}
\item 
In the case $M_1 = M_A$, comparing the formula for $m_3$ in Eq.\
(\ref{Eq:m3}) with the formulae for the washout parameters $\widetilde
m_{1,\alpha}$ in Tab.\ \ref{tab:DecayAsym} yields
\begin{equation}
\widetilde m_{1,\mu} = \widetilde m_{1,\tau} \approx \frac{1}{2} m_3 \;.
\end{equation}
Since for hierarchical light neutrino masses $m_3 \approx \sqrt{\Delta
m^2_{31}} \approx 0.05$ eV, we find that for these flavours we are
constrained to be in the strong washout regime, where $\widetilde
m_{1,\alpha} \gg m^*$ (c.f.~Fig.\ \ref{fig:eta}).  On the contrary,
$\widetilde m_{1,e}$ is significantly smaller since $|A_1| \ll
|A_2|,|A_3|$ and $\widetilde m_{1,e}$ is thus typically in the weak
(optimal) region for $\widetilde m_{1,e} \ll m^*$ ($\widetilde
m_{1,e}\approx m^*$).

Note that in the flavour-independent approximation, the washout parameter would be given by $\widetilde m_1 = \sum_\alpha \widetilde m_{1,\alpha} \approx m_3$. The fact that we expect weak (or optimal) washout for $\widetilde m_{e}$ thus provides an important difference between flavour-dependent leptogenesis compared to the flavour-independent approximation. We will analyze this situation in more detail in Sec.~\ref{A1andLSD}. 

\item In the case $M_1 = M_B$, from comparing the formula for $m_2$ in 
Eq.\ (\ref{Eq:m2}) with the formula for the washout parameter $\widetilde  m_{1,e}$ in Tab.\ \ref{tab:DecayAsym}, we infer that 
\begin{eqnarray}
\widetilde m_{1,e} \approx  s_{12}^2 m_2 \; .
\end{eqnarray}
Furthermore, from Eq.\ (\ref{Eq:t12}) and from the SD conditions  Eq.~(\ref{SDcond}) it follows that $B_2$ and/or $B_3$ are of the order of $B_1$ in order to generate large neutrino mixing $\theta_{12}$, leading to the expectations 
\begin{eqnarray}
\widetilde m_{1,\mu} \;\: \mbox{and/or} \;\: \widetilde m_{1,\tau} = {\cal O}(m_2) \; .
\end{eqnarray}

\item In the case $M_1 = M_C$, Eq.\ (\ref{Eq:m1}) implies that  
the washout parameters $\widetilde m_{1,\alpha}$ are typically $\lesssim {\cal O}(m_1)$ and therefore, with a hierarchical neutrino mass spectrum $m_1 \ll m_2 < m_3$, they are in general in the weak (or optimal) washout region. 
\end{itemize}

\noindent A summary of the above derived constraints on the washout parameters $\widetilde m_{1,\alpha}$ in the classes of models with sequential dominance is given in  Tab.~\ref{tab:FlavouredWashout}.

\begin{table} 
  \centering 
  \begin{eqnarray*} 
  \begin{array}{|c||c|c|c|} 
 \hline 
\mbox{Type of SD}
&
\mbox{washout by}\;\widetilde m_{1,e}  \vphantom{\frac{f^f}{f^f}}
&
\mbox{washout by}\;\widetilde m_{1,\mu}
&
\mbox{washout by}\;\widetilde m_{1,\tau}
\\
\hline\hline 
M_1 = M_A
&
\mbox{weak/optimal}     \vphantom{\frac{f^f}{f^f}}
&
\mbox{strong, $\widetilde m_{1,\mu} \approx \tfrac{1}{2} m_3$}
&
\mbox{strong, $\widetilde m_{1,\tau} \approx \tfrac{1}{2} m_3$}
\\
\hline
M_1 = M_B
&
\mbox{optimal, $\widetilde m_{1,e} \approx s_{12}^2 m_2$}     \vphantom{\frac{f^f}{f^f}}
& 
\lesssim {\cal O}(m_2)^*
&
\lesssim {\cal O}(m_2)^*
\\
\hline 
M_1 = M_C
&
\mbox{weak/optimal}     \vphantom{\frac{f^f}{f^f}}
&
\mbox{weak/optimal}
&
\mbox{weak/optimal}
\\
\hline 
\end{array} 
\end{eqnarray*} 
\caption{
\label{tab:FlavouredWashout} Flavour specific type of washout for the considered classes of models. (${}^*$At least one out of $\widetilde m_{1,\mu},\widetilde m_{1,\tau}$ has to be ${\cal O}(m_2)$ for $M_1 = M_B$.) } 
\end{table}

\subsubsection{Remaining Cases: Only $\boldsymbol{y_\tau}$ in Equilibrium and the Flavour-Independent Case}\label{Sec:2otherCases}
As discussed in Sec.~\ref{Sec:FML}, all three flavours are only treated differently for 
temperatures smaller than $10^9 \: \mbox{GeV}$ in the SM and larger than $(1+\tan^2 \beta)\times 10^9 \: \mbox{GeV}$ in the MSSM. 

In the temperature ranges 
$10^9 \: \mbox{GeV} \ll M_1 \ll 10^{12} \: \mbox{GeV}$ in the SM and 
$(1+\tan^2 \beta)\times 10^9 \: \mbox{GeV} \ll M_1 \ll (1+\tan^2 \beta)\times 10^{12} \: \mbox{GeV}$ in the MSSM, only the $y_\tau$ is in equilibrium and thus only the $\tau$ flavour is treated separately in the Boltzmann equations while the $e$ and $\mu$ flavours are indistinguishable. In this temperature range, one can use the combined decay asymmetry and washout parameter
\begin{eqnarray}
\varepsilon_{1,2} &\equiv& \varepsilon_{1,e}+\varepsilon_{1,\mu} \; ,\\
\widetilde m_{1,2}&\equiv& \widetilde m_{1,e} + \widetilde m_{1,\mu}\; ,
\end{eqnarray}
instead of $\varepsilon_{1,e},\varepsilon_{1,\mu}$ and $\widetilde m_{1,e},\widetilde m_{1,\mu}$ and solve the Boltzmann equations for $Y_{\Delta_2} \equiv Y_{\Delta_e + \Delta_\mu}$ and $Y_{\Delta_\tau}$.

Above $10^{12} \: \mbox{GeV}$ in the SM and 
$(1+\tan^2 \beta)\times 10^{12} \: \mbox{GeV}$ in the MSSM, all flavours are combined to the flavour-independent decay asymmetry and washout parameter 
\begin{eqnarray}
\varepsilon_{1} &\equiv& \varepsilon_{1,e}+\varepsilon_{1,\mu}+\varepsilon_{1,\tau} \; ,\\
\widetilde m_{1}&\equiv& \widetilde m_{1,e} + \widetilde m_{1,\mu} + \widetilde m_{1,\tau}
\end{eqnarray}
and one can solve the Boltzmann equations for the total asymmetry $Y_\Delta \equiv Y_{\Delta_e + \Delta_\mu + \Delta_\tau}$. 

The properties of the decay asymmetries and washout parameters in these two cases can be discussed analogously to Secs.~\ref{Sec:Decay} and \ref{Sec:Washout}. 
The results are presented in Tabs.~\ref{tab:CaseTau} and \ref{tab:CaseNoYeinEq}.

\begin{table}[t] 
  \centering 
  \begin{eqnarray*} 
  \begin{array}{|c||c|c||c|c|} 
 \hline 
\mbox{Type of SD}
&
  \varepsilon_{1,2}  \vphantom{\frac{f^f}{f^f}}
&
  \varepsilon_{1,\tau}
&
\mbox{washout by}\;\widetilde m_{1,2}  \vphantom{\frac{f^f}{f^f}}
&
\mbox{washout by}\;\widetilde m_{1,\tau}
\\
\hline\hline
M_1 = M_A  \vphantom{\frac{f^f}{f^f}}
& 
\lesssim {\cal O}(\tfrac{m_2}{m_3}) \varepsilon^{\mathrm{max}}
&
\lesssim {\cal O}(\tfrac{m_2}{m_3}) \varepsilon^{\mathrm{max}}
&
\mbox{$\widetilde m_{1,2} \approx \tfrac{1}{2} m_3$}\vphantom{\frac{f^f}{f^f}}
&
\mbox{$\widetilde m_{1,\tau} \approx \tfrac{1}{2} m_3$}
\\
\hline 
M_1 = M_B  \vphantom{\frac{f^f}{f^f}}
&
\lesssim \varepsilon^{\mathrm{max}}
&
\lesssim \varepsilon^{\mathrm{max}}
& 
\lesssim {\cal O}(m_2) \vphantom{\frac{f^f}{f^f}}
&
\lesssim {\cal O}(m_2) 
\\
\hline
M_1 = M_C  \vphantom{\frac{f^f}{f^f}}
&
\lesssim  \varepsilon^{\mathrm{max}}
&
\lesssim  \varepsilon^{\mathrm{max}}
&
\mbox{weak/optimal}     \vphantom{\frac{f^f}{f^f}}
&
\mbox{weak/optimal}
\\
\hline 
\end{array} 
\end{eqnarray*} 
\caption{
\label{tab:CaseTau} Constraints and theoretical expectations on the flavour specific decay asymmetries and the flavour specific type of washout for the case where only the tau Yukawa coupling $y_\tau$ is in thermal equilibrium. For the $e$ and $\mu$ flavours, the combined quantities $\varepsilon_{1,2} = \varepsilon_{1,e}+\varepsilon_{1,\mu}$ and $\widetilde m_{1,2} = \widetilde m_{1,e} + \widetilde m_{1,\mu}$ are considered.} 
\end{table}

\begin{table}[t] 
  \centering 
  \begin{eqnarray*} 
  \begin{array}{|c||c||c|} 
 \hline 
\mbox{Type of SD}
&
\mbox{relation for}\;  \varepsilon_{1}  \vphantom{\frac{f^f}{f^f}}
&
\mbox{washout by}\;\widetilde m_{1}  \vphantom{\frac{f^f}{f^f}}
\\
\hline\hline
M_1 = M_A  \vphantom{\frac{f^f}{f^f}}
&
\lesssim {\cal O}(\tfrac{m_2}{m_3}) \varepsilon^{\mathrm{max}}
&
\mbox{strong, $\widetilde m_{1} = m_3$}\vphantom{\frac{f^f}{f^f}}
\\
\hline 
M_1 = M_B  \vphantom{\frac{f^f}{f^f}}
&
\lesssim \varepsilon^{\mathrm{max}}
& 
\lesssim {\cal O}(m_2) \vphantom{\frac{f^f}{f^f}}
\\
\hline
M_1 = M_C  \vphantom{\frac{f^f}{f^f}}
&
\lesssim  \varepsilon^{\mathrm{max}}
&
\mbox{weak/optimal}     \vphantom{\frac{f^f}{f^f}}
\\
\hline 
\end{array} 
\end{eqnarray*} 
\caption{
\label{tab:CaseNoYeinEq} Constraints and theoretical expectations on the flavour specific decay asymmetries and the flavour specific type of washout for the case where no charged lepton Yukawa coupling is in equilibrium. All flavours are combined effectively to a single flavour in the Boltzmann equations and the relevant quantities are the flavour-independent ones, $\varepsilon_{1} = \varepsilon_{1,e} +\varepsilon_{1,\mu} +\varepsilon_{1,\tau}$ and $\widetilde m_{1}\equiv \widetilde m_{1,e} + \widetilde m_{1,\mu} + \widetilde m_{1,\tau}$.} 
\end{table}

\subsection{Leptonic CP Violation and the MNS - Leptogenesis Link}\label{Sec:MNSLGlink}
In \cite{King:2002qh} it was shown how, in the case of SD where the
lightest right-handed neutrino was the dominant one, there was a link
between the MNS Dirac CP violating phase $\delta$ which appears in
neutrino oscillations and the phase which is relevant for
leptogenesis.  However the analysis in \cite{King:2002qh} was based on
the flavour-independent formulation of leptogenesis, and it is
therefore interesting to revisit this analysis in the light of
flavour-dependent leptogenesis.

Eqs.~(\ref{tanetadelta}) and (\ref{eta23}) show explicitly that  
the MNS Dirac CP violating phase $\delta$ is a function
of only two seesaw phases $\zeta_2,\ \zeta_3$,
as in \cite{King:2002qh}.
Since these are the only remaining seesaw phases it must also
be the case that leptogenesis must depend also on these
same two phases. This opens up the possibility of a link
between the leptogenesis phase and the oscillation MNS phase $\delta$,
as in \cite{King:2002qh}. In the flavour-independent treatment
of leptogenesis in \cite{King:2002qh} the relevant leptogenesis
phase was $\zeta$ and the leptogenesis-MNS link was 
given by Eqs.~(\ref{taneta}), (\ref{tanetadelta}). 
However, in the present case, flavour-dependent
leptogenesis will depend on both phases
$\zeta_2,\ \zeta_3$, as we now discuss.

\begin{itemize}
\item
For the case $M_1=M_A$, assuming that $A_1=0$,
we have 
\begin{eqnarray}
\varepsilon_{1,\alpha} 
\approx -\frac{3 M_1}{16 \pi } 
\frac{\mathrm{Im}\left[ A_\alpha^* B_\alpha  (A^\dagger B) \right]}{ M_B\,(A^\dagger A)}\;.
\end{eqnarray}
From this result it is clear that 
\bea
\varepsilon_{1,e} & = & 0,\nonumber \\ 
\varepsilon_{1,\mu} & \propto & 
+|A_2||B_2|\sin (\zeta + \zeta_2),\nonumber \\ 
\varepsilon_{1,\tau} & \propto & 
+|A_3||B_3|\sin (\zeta + \zeta_3).
\eea
From the previous section we see that 
\beq
Y_B \propto Y_{\Delta_\mu} + Y_{\Delta_\tau}
\eeq
with $Y_{\alpha}$ given in Eq.~(\ref{Eq:eta_aa}).
Since the efficiency factor $\eta_{\mu}$ depends on 
$\tilde{m}_{1,\mu}\approx |A_2|^2v^2/M_A$
and 
$\eta_{\tau}$ depends on 
$\tilde{m}_{1,\tau}\approx |A_3|^2v^2/M_A$,
with $|A_2|\approx |A_3|$ we see that,
$\eta_{\mu}\approx \eta_{\tau}$,
and
\beq
Y_B\propto |B_2|\sin (\zeta + \zeta_2)
+ |B_3|\sin (\zeta + \zeta_3),
\label{YB1}
\eeq
which shows explicitly how leptogenesis depends on the
two phases $\zeta_2,\ \zeta_3$ in this case.

\item
For the case $M_1=M_B$, assuming that $A_1=0$,
we have 
\begin{eqnarray}
\varepsilon_{1,\alpha} 
\approx -\frac{3 M_1}{16 \pi } 
\frac{\mathrm{Im}\left[ B_\alpha^* A_\alpha  (B^\dagger A) \right]}{ M_B\,(B^\dagger B)}\;.
\end{eqnarray}
From this result it is clear that 
\bea
\varepsilon_{1,e} & = & 0,\nonumber \\ 
\varepsilon_{1,\mu} & \propto & 
-|A_2||B_2|\sin (\zeta + \zeta_2),\nonumber \\ 
\varepsilon_{1,\tau} & \propto & 
-|A_3||B_3|\sin (\zeta + \zeta_3).
\eea
Since now $\eta_{\mu}$ depends on the washout parameter  
$\tilde{m}_{1,\mu}\approx |B_2|^2 v^2/M_B$
and 
$\eta_{\tau}$ depends on 
$\tilde{m}_{1,\tau}\approx |B_3|^2 v^2/M_B$,
the two efficiencies are in general unequal in this case,
and with $|A_2|\approx |A_3|$ we see that,
\beq
Y_B\propto -\eta_{\mu}|B_2|\sin (\zeta + \zeta_2)
-\eta_{\tau} |B_3|\sin (\zeta + \zeta_3),
\label{YB2}
\eeq
which shows explicitly how leptogenesis depends on the
two phases $\zeta_2,\ \zeta_3$ in this case.

\item
Clearly, for the case $M_1=M_C$, there is no leptogenesis-MNS
link since leptogenesis will depend on phases associated
with the right-handed neutrino of mass $M_C$, which does not
contribute significantly to the seesaw mechanism, and therefore
these phases will not significantly contribute to the MNS phases.

\end{itemize}

The leptogenesis-MNS link in the flavour-dependent case is therefore
rather similar to the link in the flavour-independent case,
namely the connection between leptogenesis and $\delta$
is made via the remaining two seesaw phases $\zeta_2,\ \zeta_3$ which are 
responsible for both leptogenesis and $\delta$.
These two phases fix $\delta$ as in 
Eqs.~(\ref{tanetadelta}), (\ref{eta23}), 
which are valid for all types of SD,
including the cases $M_1=M_A$ and $M_1=M_B$.
However the two phases $\zeta_2,\ \zeta_3$ contribute to 
leptogenesis differently for the cases $M_1=M_A$ and $M_1=M_B$,
as shown in Eqs.~(\ref{YB1}) and (\ref{YB2}), where both these results
differ from the flavour-independent result in which leptogenesis
is a simple function of $\zeta$.

In order to obtain a more precise leptogenesis-MNS link,
one can reduce the number of phases still further by assuming
an additional zero Yukawa coupling, in addition to assuming that
$A_1=0$. For example if we additionally assume that $B_3=0$, 
which implies that $\varepsilon_{1,\tau}=0$, 
then this removes the phase $\zeta_3$, and we find that
$\zeta = \zeta_2$.
From Eqs.~(\ref{real12}) - (\ref{eta}), we obtain for the 
MNS phase $\delta$ \cite{King:2002qh}: 
\begin{eqnarray}
\delta = - 2 \zeta \; .\label{deltaeta2}
\end{eqnarray}
Under the same assumption that $B_3=0$ (as well as $A_1=0$),
for the case $M_1=M_A$ we see that 
\beq
Y_B \propto +\sin (2\zeta ),
\label{YB11}
\eeq
while for the case $M_1=M_B$ we see that 
\beq
Y_B \propto -\sin (2\zeta ),
\label{YB21}
\eeq
where in both cases leptogenesis depends on a single
phase $\zeta$, which is directly related to the 
MNS phase $\delta$ in Eq.~(\ref{deltaeta2}). 
The sign of the CP violating phase $\delta$ measurable in 
high precision neutrino oscillation experiments, or more precisely the sign of 
$\sin(\delta)$, therefore will
depend on whether $M_1=M_A$ or $M_1=M_B$.
If $M_1=M_A$ then $\sin(\delta)$ must be negative since baryon asymmetry
of the universe in Eq.~(\ref{YB11}) must be positive.
If $M_1=M_B$ then $\sin(\delta)$ must be positive since baryon asymmetry
of the universe in Eq.~(\ref{YB21}) must be positive.
Thus the measured sign of the oscillation phase $\delta$ 
is capable of distinguishing between the 
two types of SD, namely $M_1=M_A$ or $M_1=M_B$,
under the assumption of the two texture zeroes $A_1 = 0$ and $B_3 = 0$.\footnote{A different possible choice of two texture zeros, which can be discussed analogously, is $A_1 = B_2 = 0$. In this case, the relation between $\delta$ and $\zeta$ would be given by $\delta = - 2 \zeta + \pi$, leading to the prediction of the opposite sign of $\sin(\delta)$ from the observed baryon asymmetry.}

\subsection{Examples}\label{examples}

\subsubsection{The Role of $A_1$ in the Case $M_1 = M_A$}\label{A1andLSD}
As we have remarked in the discussion of the flavour specific washout
factors $\widetilde m_{1,\alpha}$, one particularly interesting
difference between flavour-dependent treatment and flavour-independent
approximation is the possibility to have an optimal washout parameter
$\widetilde m_{1,e}$ and strong washout in the flavours $\mu$ and
$\tau$ since in the flavour-independent 
treatment this leads to a strong total washout.  
We will therefore discuss the role of $A_1$ in the
case $M_1 = M_A$, sometimes also called light sequential dominance
(LSD), in more detail.

Let us start with the special case is $A_1 = 0$, where due to $|A_2| \approx |A_3|$ required by $\theta_{23} \approx 45^\circ$ (c.f.\ Eq.~(\ref{Eq:t23})) the 
washout parameters satisfy
\begin{eqnarray}
\widetilde m_{1,\mu} \approx \widetilde m_{1,\tau} \approx \frac{1}{2} m_3\; , \;\,
\mbox{leading to} \; \eta_{\mu} \approx \eta_{\tau}\approx 
\eta \left(A_{\tau\tau}\frac{1}{2} \frac{m_3}{m^*},\frac{m_3}{m^*}\right).
\end{eqnarray}
We can now estimate for the baryon-to-entropy density in the MSSM, using Eqs.~(\ref{Eq:eta_aa_MSSM}) and (\ref{Eq:YB3f}),
\begin{eqnarray}
Y_B \approx - \frac{10}{31}\,\eta\left(A_{\tau\tau}\frac{1}{2} \frac{m_3}{m^*},\frac{m_3}{m^*}\right) \, ( \varepsilon_{1,\mu} + \varepsilon_{1,\tau}) \, 
\left[ Y^{\mathrm{eq}}_{N_1}(z\gg1) + Y^{\mathrm{eq}}_{\widetilde N_1}(z\gg1)\right] .
\end{eqnarray} 
Apart from minor modifications in the efficiency factor, the
flavour-dependent treatment is thus similar to the flavour-independent
approximation in this case \cite{Hirsch:2001dg}.
   
\begin{figure}
 \centering
 \includegraphics[scale=1.0,angle=0]{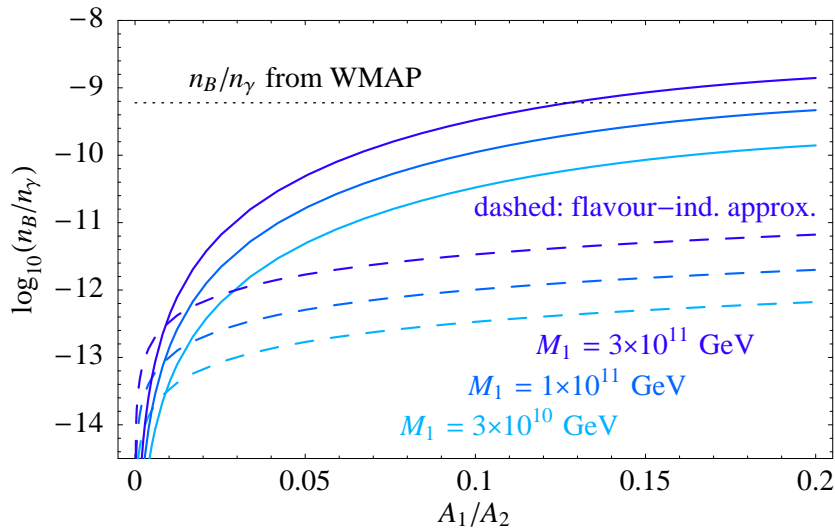}
 \caption{\label{fig:A1}
Example in the MSSM (with large $\tan \beta = 50$), which illustrates the difference between flavour-dependent treatment and  flavour-independent approximation for the case $M_1 = M_A$ if $A_1$ is non-zero. We have chosen $\phi_{A_1} = - \tfrac{\pi}{2}, \phi_{B_3} = \pi, \phi_{A_2} = \phi_{A_3} = \phi_{B_3} = \phi_{C_i} = 0, B_3 = 0, C_i = 0$. The other parameters are determined from the experimental data assuming SD. For the parameters under consideration, $\theta_{13}$ remains below the present experimental upper bound. 
 }
\end{figure}

The importance of a flavour-dependent treatment becomes clear if we
allow for a non-zero $A_1 \ll A_2 \approx A_3$. As we have discussed
above, $\widetilde m_{1,e}$ can now be optimal whereas $\widetilde
m_{1,\mu}$ and $\widetilde m_{1,\tau}$ correspond to a strong
washout. In the flavour-independent approximation, the total washout
parameter $\widetilde m_1 = \sum_\alpha \widetilde m_{1,\alpha}
\approx m_3$ implies strong washout.
In Fig.~\ref{fig:A1}, the flavour-dependent treatment is compared to the flavour-independent approximation in an example with non-zero $A_1$. Dashed lines correspond to the flavour-independent approximation and solid lines stand for the same examples with  
flavour-effects included. 
For non-zero $A_1/A_2$ above roughly $0.05$, the produced baryon asymmetry is significantly enhanced in the flavour-dependent case (more than two orders of magnitude in this example for $A_1/A_2 = 0.2$) because this induces an efficiency in the $e$-flavour which is close to optimal. 
In the flavour-independent approximation, the common efficiency factor  $\eta^\mathrm{ind}(\widetilde m_1/m^*)$ is much lower due to the strong washout. 

\vspace{7mm}

\subsubsection{Leptogenesis and Tri-Bimaximal Mixing}
As discussed in Sec.~\ref{SD}, tri-bimaximal mixing can be realized in
SD in a natural way via vacuum alignment and is sometimes referred to
as constrained sequential dominance (CSD) \cite{King:2005bj}. 
As can be seen directly
from Eq.~(\ref{zero}), which states that $A^\dagger B = B^\dagger A =
0$, together with the formulae for the decay asymmetries
$\varepsilon_{1,\alpha}$ (c.f.\ Tab.~\ref{tab:DecayAsym}), in CSD the MSSM 
decay asymmetries reduce to
\begin{eqnarray}
\varepsilon_{1,\alpha} &=& -\frac{3 M_1}{8 \pi } 
\frac{\mathrm{Im}\left[ A_\alpha^* C_\alpha  (A^\dagger C) \right]}{ M_C\,(A^\dagger A)}
\;\lesssim\; \varepsilon^{\mathrm{max}}  \, 
{\cal O}\!\left(\frac{m_1}{m_3}\right)
\quad \mbox{for $M_1 = M_A$, and} \label{epsCSDmm}\\
\varepsilon_{1,\alpha} &=& -\frac{3 M_1}{8 \pi } 
\frac{\mathrm{Im}\left[ B_\alpha^* C_\alpha  (B^\dagger C) \right]}{ M_C\,(B^\dagger B)}
\;\lesssim\; \varepsilon^{\mathrm{max}}  \, 
{\cal O}\!\left(\frac{m_1}{m_3}\right)
\quad \mbox{for $M_1 = M_B$.} \label{epsCSDtt}
\end{eqnarray}
For $M_1 = M_A$ and $M_1 = M_B$, the decay asymmetries are thus suppressed by $m_1/m_3$.

We would like to remark at this point that in realistic models, tri-bimaximal mixing is typically realized for the neutrino mass matrix in a basis where the charged lepton mass matrix is not entirely diagonal. However, including such charged lepton corrections do not change the above conclusions since $A^\dagger B = B^\dagger A = 0$ remains unchanged by a change to the basis where the charged lepton mass matrix is diagonal, even though the lepton mixing is no longer exactly tri-bimaximal. In this respect, the conclusions are the same in leptogenesis with correct flavour-dependent treatment and in the flavour-independent approximation.\footnote{Note that in the R-matrix parameterization with $R=\mathbbm{1}$, where all mixings and phases in $\lambda_\nu$ are introduced via the leptonic mixing matrix $U_\mathrm{MNS}$, all flavour-specific decay asymmetries also vanish exactly as in the flavour-independent, since they are proportional to $(\lambda_\nu^\dagger \lambda_\nu)_{1J}$ which is zero for $1 \not= J$.}   
In general, however, we stress that charged lepton corrections \cite{Antusch:2005kw} do have an effect on leptogenesis in the flavour-dependent treatment, in contrast to the flavour-independent approximation.  
 
In contrast to the cases $M_1 = M_A$ and $M_1 = M_B$, tri-bimaximal mixing via CSD has excellent prospects for leptogenesis in the classes of models with 
$M_1 = M_C$, as we now discuss.

\subsubsection{Tri-Bimaximal Mixing and CP phases in the Case $M_1=M_C$} 
\label{Sec:CSDinMC}
To be explicit, let us consider the specific choice $A=(0,a,-a)$,
$B=(b,b,b)$ with real $a$ and $b$ and furthermore a texture zero in
the (1,1)-element of $\lambda_\nu$, i.e. $C$ of the form
$C=(0,C_2,C_3)$. Then, $\varepsilon_{1,e}=0$ and the remaining two
decay asymmetries in the MSSM simplify to
\begin{eqnarray}
\varepsilon_{1,\mu} &=& -\frac{3 M_1}{8 \pi } \left\{
\frac{\mathrm{Im}\left[ C_2^* a^2 (C_2^* - C_3^*) \right]}{ M_A (C^\dagger C)} 
+
\frac{\mathrm{Im}\left[ C_2^* b^2 (C_2^* + C_3^*) \right]}{ M_B (C^\dagger C)}
\right\} \nonumber \\
&\approx&
-\frac{3 M_1}{8 \pi v^2} \left\{
\frac{\mathrm{Im}\left[ C_2^* (C_2^* - C_3^*) \right]}{ (C^\dagger C)} 
\, \frac{m_3}{2}
+
\frac{\mathrm{Im}\left[ C_2^* (C_2^* + C_3^*) \right]}{ (C^\dagger C)} 
\,s_{12}^2 m_2
\right\}, \label{eq:emmCSDMC}\\
\varepsilon_{1,\tau} &=& -\frac{3 M_1}{8 \pi } \left\{
\frac{\mathrm{Im}\left[- C_3^* a^2 (C_2^* - C_3^*) \right]}{ M_A (C^\dagger C)} 
+
\frac{\mathrm{Im}\left[ C_3^* b^2 (C_2^* + C_3^*) \right]}{ M_B (C^\dagger C)}
\right\} \nonumber \\
&=&
-\frac{3 M_1}{8 \pi v^2} \left\{
\frac{\mathrm{Im}\left[- C_3^* (C_2^* - C_3^*) \right]}{ (C^\dagger C)} 
\, \frac{m_3}{2} 
+
\frac{\mathrm{Im}\left[ C_3^* (C_2^* + C_3^*) \right]}{ (C^\dagger C)} 
\,s_{12}^2 m_2
\right\}\label{eq:ettCSDMC}.
\end{eqnarray}  
One can see that the two terms from the dominant (subdominant) RH neutrinos can easily  contribute $\approx \varepsilon^\mathrm{max}$ ($\approx \varepsilon^\mathrm{max} s_{12}^2 m_2/m_3$) to the decay asymmetries. This is nearly the optimal case for leptogenesis with hierarchical neutrinos. Furthermore, the washout parameters $\widetilde m_{1,\alpha}$ are ${\cal O}(m_1)$, which allows optimal efficiencies $\eta_{\alpha}$ if $m_1 \approx m^* \approx 10^{-3}$ eV.  

Let us now discuss under which conditions flavour effects are
important, or play only a minor role. Similar to the discussion in
Sec.~\ref{A1andLSD}, when the relevant washout parameters are equal,
as this would in our example be the case for $|C_2| = |C_3|$, the
situation is similar to the flavour-independent approximation. On the
other hand, it is clear that if the washout parameters differ for the
$\mu$ and $\tau$ flavour, the produced asymmetry in the
flavour-dependent treatment will differ in general from the
flavour-independent approximation.

\begin{figure}[t]
 \centering
 \includegraphics[scale=1,angle=0]{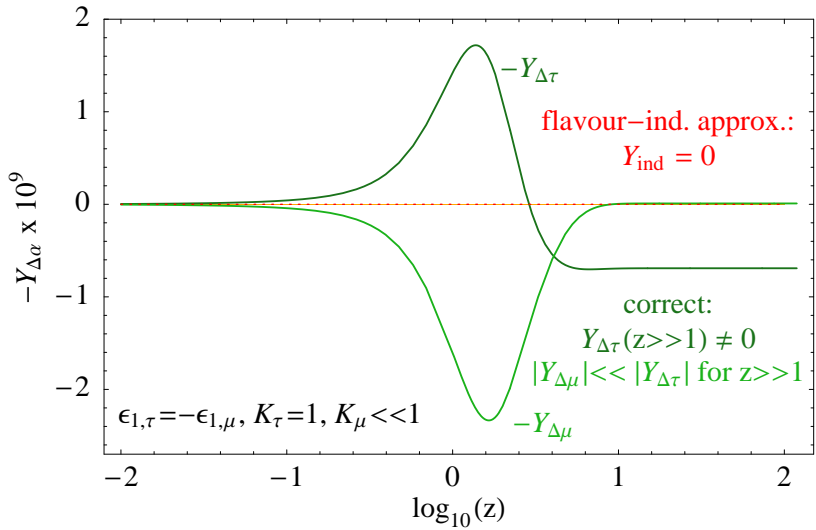}
 \caption{\label{fig:CSD}
Evolution of the flavour specific $B/3 - L_\alpha$ asymmetries $Y_{\Delta_\mu}$ and $Y_{\Delta_\tau}$ as a function of $z=M_1/T$ for the example of tri-bimaximal mixing via CSD, with $M_1 = M_C$ and the choice $C_1 = 0$. 
Complex phases $\phi_{C_2}$ and $\phi_{C_3}$ are chosen as in Eq.~(\ref{Eq:phasesCSD}), leading to $\varepsilon_{1,\tau} = - \varepsilon_{1,\mu}$. 
While the flavour-independent approximation yields zero lepton asymmetry $Y_\mathrm{ind}$, the two flavour-specific asymmetries evolve differently for different flavour-specific washout parameters $K_\mu = \widetilde m_{1,\mu}/m^*$ and $K_\tau = \widetilde m_{1,\tau}/m^*$, allowing a final baryon asymmetry to be produced (c.f.\ Eq.~(\ref{Eq:CSDexample})). 
This situation can be realized, for example, with  
$M_1 = M_C = 1.85 \times 10^{11}$ GeV, 
$|C_1|=0, 
|C_2|=1.69 \times 10^{-4}, |C_3|=3.09 \times 10^{-3},
\phi_{C_2}= - \pi/2, \phi_{C_3} = 0.028$ and
$M_A = 5 \times 10^{12}$ GeV, 
$|A_1|=0,|A_2|=|A_3|=0.063, \phi_{A_2}=0, \phi_{A_3}=0$. 
The subdominant SD contribution to the decay asymmetries vanishes for $B_2 = 0, \phi_{B_3} =  \phi_{C_3}$. 
The decay asymmetries are given by $\eps_{1,\tau} \approx - 10^{-6}, \eps_{1,\mu} \approx 10^{-6}$, and the washout parameters are
$K_\tau \approx 1, K_\mu \approx 3 \times 10^{-3}$. 
}
\end{figure}

In addition, the values of the decay asymmetries depend strongly on the complex phases, in our simplified example on the two phases $\phi_{C_2}$ and $\phi_{C_3}$ of $C_2$ and $C_3$, respectively. If $|C_2| \not= |C_3|$ and $\phi_{C_2} \not= \phi_{C_3}$, the differences between flavour-dependent treatment and flavour-independent approximation can be even more dramatic: 
To illustrate this, let us focus on the dominant terms proportional to $m_3$ in $\varepsilon_{1,\mu}$ and $\varepsilon_{1,\tau}$ given in Eqs.~(\ref{eq:emmCSDMC}) and (\ref{eq:ettCSDMC}), which are proportional to 
\begin{eqnarray}
\varepsilon_{1,\mu} &\propto& \mathrm{Im}\left[ C_2^* (C_2^* - C_3^*) \right], \\
\varepsilon_{1,\tau} &\propto& \mathrm{Im}\left[- C_3^* (C_2^* - C_3^*) \right].
\end{eqnarray}
On the other hand, in the flavour-independent approximation, the leading contribution to the decay asymmetry $\varepsilon_1 = \sum_\alpha \varepsilon_{1,\alpha} = \varepsilon_{1,\mu}+ \varepsilon_{1,\tau}$ is proportional to 
\begin{eqnarray}
\varepsilon_1 
&\propto& \mathrm{Im}\left[ (C_2^* - C_3^*)^2\right]. 
\end{eqnarray}
We can now imagine the situation that $\mathrm{Im}\left[ (C_2^* - C_3^*)^2\right]$ vanishes exactly, which would correspond to a choice of phases $\phi_{C_2}$ and $\phi_{C_3}$ such that $\mbox{Arg}(C_2 - C_3) \in \{0,\tfrac{\pi}{2}\}$ (mod $\pi$). 
Obviously, the phases $\phi_{C_2}$ and $\phi_{C_3}$ can be chosen to satisfy this (for given $|C_2| \not= |C_3|$) also with non-trivial $\phi_{C_2},\phi_{C_3} \notin \{0,\tfrac{\pi}{2}\}$, for instance with 
\begin{eqnarray}\label{Eq:phasesCSD}
\frac{\sin(\phi_{C_2})}{\sin(\phi_{C_3})} = \frac{|C_3|}{|C_2|},
\end{eqnarray} 
which implies $\mathrm{Im}\left[(C_2^* - C_3^*)\right]=0$.
Then, we have a situation where 
\begin{equation}
\varepsilon_{1,\mu} = - \varepsilon_{1,\tau} =: \,\hat \varepsilon
\end{equation}
 are non-zero, whereas the corresponding (leading) contribution to $\varepsilon_1$ vanishes exactly. Since the washout parameters are different, the leading contribution to the decay asymmetries, using Eqs.~(\ref{Eq:eta_aa_MSSM}) and (\ref{Eq:YB3f}) in the MSSM, can thus generates a baryon asymmetry of
\begin{eqnarray}\label{Eq:CSDexample}
Y_B \approx - \frac{10}{31}\, \,(\eta_{\mu} - \eta_{\tau}) \,\, \hat \varepsilon \, \left[ Y^{\mathrm{eq}}_{N_1}(z\gg1) + Y^{\mathrm{eq}}_{\widetilde N_1}(z\gg1)\right]\;, 
\end{eqnarray} 
 proportional to the difference of the efficiencies $\eta_{\mu}=\eta(A_{\mu\mu}\widetilde m_{1,\mu}/m^*,(\sum_\alpha \widetilde m_{1,\alpha})/m^*)$ and $\eta_{\tau}=\eta(A_{\tau\tau}\widetilde m_{1,\tau}/m^*,(\sum_\alpha \widetilde m_{1,\alpha})/m^*)$.
Let us consider the MSSM with $\tan \beta = 50$ as an example, such that all flavours are to be treated independently for temperatures below about $2.5 \times 10^{12}$ GeV. 
The flavour-specific washout parameters $\widetilde m_{1,\mu}$ and $\widetilde m_{1,\tau}$ are given by $C_2^2 v^2_\mathrm{u}/M_1$ and $C_3^2 v^2_\mathrm{u}/M_1$, respectively (c.f.\ Tab.~\ref{tab:DecayAsym}). For instance, $\widetilde m_{1,\tau} \approx m^*$ ($K_\tau \approx 1$) could lead to a nearly optimal efficiency factor $\eta_{\tau}$, while $C_2 \ll C_3$, and correspondingly $\widetilde m_{1,\mu}\ll m^*$ ($K_\mu \ll 1$), could imply a very low efficiency $\eta_\mu \ll \eta_{\tau}$ (c.f.\ Fig.~\ref{fig:eta}). 
According to Eq.~(\ref{Eq:CSDexample}), in the correct flavour-dependent treatment a baryon asymmetry can be generated, while in the flavour-independent approximation (in leading order in SD) it would vanish. This situation is illustrated in Fig.~\ref{fig:CSD}.

\section{Summary and Conclusions}\label{concl}
We have studied thermal leptogenesis in a class of neutrino mass models
based on the see-saw mechanism and SD, taking into account lepton
flavour in the Boltzmann equations in both the SM and MSSM.
SD models are a very well motivated class of models which have been
widely applied to unified flavour models, and thus are quite
representative of the general class of three right-handed neutrino
models where the importance of flavour-dependent leptogenesis
effects has hitherto not been evaluated.  

In general, flavour-dependent effects of leptogenesis are relevant 
at low enough temperatures (set by the lightest right-handed neutrino mass)
such that at least one charged lepton flavour is in thermal equilibrium. 
When this condition is
met, flavour-dependent effects are important when the efficiency
factors $\eta_{\alpha}$ differ significantly for the distinguishable
flavours. The efficiency factors $\eta_{\alpha}$ depend
on $\tilde{m}_{1,\alpha}$ which in SD are determined from the
magnitudes of the Yukawa couplings of the lightest right-handed
neutrino to the different lepton flavours, as indicated in Tab.~\ref{tab:DecayAsym}. 

If the dominant right-handed neutrino is
the lightest one, then only $\eta_{e}$ may differ,
since $|A_2|\approx |A_3|$ implies that $\eta_{\mu}\approx \eta_{\tau}$, 
and flavour-dependent effects are only important when $A_1\neq 0$,
corresponding to non-zero texture element in
the (1,1)-element of the neutrino Yukawa matrix $\lambda_\nu$. We have studied this case
numerically and shown that dramatic effects can result.
Moreover, we have also seen that, even with $A_1=0$,
flavour-dependent effects are relevant for the leptogenesis-MNS
link since the decay asymmetries $\epsilon_{1,\mu},\epsilon_{1,\tau}$
may differ due to $B_2\neq B_3$. 
An extreme example of this is when there are two zero textures in the dominant and leading subdominant parts of $\lambda_\nu$, for instance $A_1=B_3=0$, in which case the cosmological CP violating phase for leptogenesis is directly related to the CP violating phase $\delta$ of the MNS matrix observable in neutrino oscillations.
Furthermore, the baryon asymmetry of the universe under this conditions determines the sign of the MNS phase $\delta$ (i.e.\ the sign of $\sin(\delta)$) for the case where the lightest right-handed neutrino is the dominant one. We have shown that this phase
is opposite to the case where the lightest right-handed neutrino
is the leading subdominant one, enabling the two types of SD to 
be distinguished experimentally.

For the case that the lightest right-handed
neutrino is the almost decoupled one, there is no leptogenesis-MNS link.
Nevertheless this case is amongst the most promising for thermal
leptogenesis since the efficiencies may be optimal.
Although this is also true in the flavour-independent treatment,
flavour-dependent effects are expected to play 
an important role when the efficiences are unequal,
corresponding to the magnitudes of the Yukawa couplings $|C_{\alpha}|$
being unequal. Unlike the previous case,
this can be achieved with a zero texture in
the (1,1)-element of $\lambda_\nu$ corresponding to 
$C_1=0$, since here we can have $|C_2|\neq |C_3|$ since 
these
couplings are unconstrained by neutrino phenomenology,
and so in general may be quite different.
Thus very large flavour-dependent effects can be present
in this case, an extreme example being when 
$\epsilon_{1,\mu}$ and  $\epsilon_{1,\tau}$ 
are equal and opposite which would lead to a zero 
result in the flavour-independent case, but a 
non-zero result in the flavour-dependent case due to the
different efficiency factors.

In conclusion, we have seen that in many 
cases flavour-dependent effects may be important for 
leptogenesis in SD models involving three right-handed neutrinos
due to the efficiencies of the distinguishable flavours being different.
The effects range from factors of three, up to differences
of a few orders of magnitude, or in extreme cases having a large non-zero 
flavour-dependent result for cases where the flavour-independent
result gives zero. In other cases 
the flavour-dependence of leptogenesis is not important
for the asymmetry, due to the efficiencies of distinguishable
flavours being equal, as in the case when the lightest right-handed
neutrino dominates the see-saw mechanism, and there is a 
zero texture in the (1,1)-element of $\lambda_\nu$.
However even in this case the relationship
between leptogenesis phases and the MNS phase is affected
by flavour-dependent effects due to the lepton asymmetries
being unequal. We conclude that flavour-dependent effects 
cannot be ignored when dealing with three right-handed neutrino
models.

\section*{Acknowledgements}
The work of S.~Antusch was supported by the EU 6$^\text{th}$
Framework Program MRTN-CT-2004-503369 ``The Quest for Unification:
Theory Confronts Experiment''. S.~F.~King would like to acknowledge 
a CERN Scientific Associateship.

\providecommand{\bysame}{\leavevmode\hbox to3em{\hrulefill}\thinspace}

\end{document}